\documentclass{aa}
\usepackage{graphicx}
\usepackage{lipsum}
\usepackage{natbib}
\usepackage{amsmath}
\usepackage{multicol}
\usepackage{stfloats}
\usepackage{color}
\usepackage{yfonts}
\usepackage{ulem}
\usepackage{xcolor}
\usepackage{txfonts}
\usepackage{booktabs}
\usepackage[colorlinks=false]{hyperref}
\begin{document} 

\title{Homogeneous search for spot transits in Kepler and TESS photometry of K -- M-type main-sequence stars}
\titlerunning{Spot transits in Kepler and TESS K -- M-dwarf light curves}

   \subtitle{}

   \author{A.~Haris
          \inst{1}
          \and
          M.~Tuomi\inst{1}
          \and 
          T.~Hackman\inst{1}
          }

   \institute{            Department of Physics, P.O.~Box 64, 00014 University of Helsinki, Finland\\
              \email{andras.haris-kiss@helsinki.fi}
             }

   \date{
   }

 
  \abstract
   {Late-type stars are known to host numerous exoplanets, and their photometric variability, primarily caused by rotational modulation, provides a unique opportunity to study starspots. As exoplanets transit in front of their host stars, they may occult darker, spotted regions on the stellar surfaces. The monitoring of starspots from planetary transits, known as transit mapping, offers a possibility to detect small dark regions on magnetically active, late-type stars. These spots may be so small that they would be undetectable to other methods used to reconstruct stellar magnetic activity.
   
   We describe a Bayesian analysis framework on the transit light curves of planets orbiting K- and M-type main-sequence stars in search for spot occultation event candidates. We present a systematic analysis of high-precision, high-cadence light curves from Kepler and TESS to detect and characterise starspots during exoplanetary transits. According to our tests, the set of criteria applied in the analysis is robust and not prone to false positives. Our sample comprises K- and M-dwarfs hosting transiting exoplanets observed by the Kepler or TESS space telescopes at a high cadence, totalling 99 planets meeting our selection criteria.
   
   After analysing $3273$ transit light curves from 99 planets, we find 105 candidates for starspot occultation events by six planets. We report new spot occultation candidates for the K~dwarfs HD~189733 and TOI-1268.
   The identified dark regions have a lower limit for radii between 1.6~degrees and 29.5~degrees and contrasts up to 0.69.
   We estimate a spot detection frequency of $3.7\%$ and $4.2\%$ for K- and M-dwarfs by TESS, and $37.5\%$ for K-dwarfs by Kepler.
   }


   \maketitle
%

\section{Introduction}

Late-type stars are the most abundant stellar objects in the Milky Way. The coolest, M-type stars make up $70\%$ of stars in our galaxy \citep{henry2006}. These stars have an abundance of short period planets, which allows for easy detection with the transit method \citep{hsu2020}. The strongest periodic signals in the light curves of these stars are caused by the common effect of starspots and stellar rotation. Spots may change the brightness of stars to a significantly higher extent than the transits of planets. To detect Earth-sized planets around late-type stars, it is essential to model stellar magnetic activity and to remove its signals from observations.

Starspots on late-type stars are typically resolved with inversion methods. These methods aim to reconstruct the star's surface brightness or magnetic field distribution by searching for model solutions that best fit the time-series observations of the star. Such methods include Doppler imaging \citep[see e.g.,][]{piskunov2002}, or photometric inversion \citep{budding1977}. However, both methods are only reliable at reconstructing the large-scale dark regions of stars, resulting in a loss of information on smaller-scale surface features.

The signs of magnetic activity and the influence of dark spots over longer timescales are expected to increase with decreasing stellar effective temperature \citep[see e.g.,][]{boro2018}. Therefore, planets of cooler, K- and M-type stars should, in theory, show more spot occultations than planets of warmer stars with a convective surface. 

Over the past fifteen years, the space-based Kepler \citep{borucki2010} and Transiting Exoplanet Survey Satellite \citep[TESS][]{ricker2015} have observed hundreds of thousands of stars and revolutionised the field of exoplanets. These instruments allow for the continuous monitoring of a large set of stars, providing high-precision light curves at a high cadence. Kepler and TESS have discovered exoplanets with the transit method, i.e., by measuring periodic dimmings of stars caused by exoplanets transiting in front of them and thereby occulting parts of their bright disks.

The light curves from these instruments are, in some cases, so precise that signatures from large spot groups on the stellar surface may become detectable when a planet passes in front of them. As the planet transits in front of a darker region on the stellar surface, the transit light curve shows a slight brightness increase above the transit baseline. The characterisation of stellar surface features based on photometric data obtained during the planetary transit is known as transit mapping \citep{silva2003}.
For a more complete review of the development of transit mapping, see \citet{baluev2021}. The term ``spot'' in the context of transit mapping can be somewhat misleading, as the observed transit signal may result from a planet crossing multiple spots or even entire active regions. In this paper, we use the terms ``spot'' and  ``dark region'' interchangeably. Thus, these terms account for the whole transited region containing starspots.

Since space-based observations of exoplanetary transits began, spot occultation events have been reported for several stars. For instance, spot occultations were reported for HD~189733 \citep{pont2007} from observations with the Hubble Space Telescope, CoRoT-2 \citep{silva2010} from observations with CoRoT \citep{baglin2006}, HAT-P-11 \citep{sanchis2011}, Kepler-45 \citep{zaleski2020}, and Kepler-411 \citep{araujo2021} from Kepler observations, and TOI-3884 \citep{almenara2022} from TESS measurements. The spot models and the criteria for detection vary greatly, which makes the comparison of results and the analysis spot populations challenging.

We aim to analyse all transit light curves of planets around K- and M-type main sequence stars observed by Kepler or TESS, using a uniform set of criteria in search for spot occultation events. This is achieved through applying a homogeneous set of criteria to assess the significance of the photometric signal of the candidate spot eclipses. Through a uniform analysis of a large sample of transit light curves, it is possible to estimate the frequency of starspot occultation events by exoplanets.

\section{Observations}

\subsection{Sample}

We selected our sample of targets to maximise the chances of spot eclipse detections. We restricted the sample to the two coolest spectral classes, K and M. These spectral classes are the ones with strongest predicted magnetic activity and highest likelihood of spot eclipses, which is favourable for spot occultation detection. We chose to use observations from the Kepler and TESS space telescopes, as these instruments provide high-precision light curves for a large number of objects. The observations from these two telescopes have comparable noise properties and have provided thousands of exoplanet detections. Kepler and TESS are thus ideal instruments for a uniform analysis of large quantities of transit light curves.

We chose the observations such that they have high cadence, as most spot eclipses are expected to happen on the timescale of less than $30$ minutes. We also required that the transit observations are obtained with a signal-to-noise ratio of at least two, to avoid wasting time searching for spot eclipses where they would most likely be undetectable.

To create our sample, we accessed the catalogue of confirmed exoplanets from the NASA Exoplanet archive\footnote{\url{https://exoplanetarchive.ipac.caltech.edu/}, accessed on July 11, 2023.} \citep{akeson2013}. Using this catalogue, we compiled our sample by choosing planets such that they:
\begin{itemize}
    \item transit a K- or M-dwarf star,
    \item have transits observed by Kepler or TESS, with a cadence of at most 200 s,
    \item have a transit depth $\delta$, that is at least two times greater than the standard deviation of fluxes $\sigma_{\mathrm{flux}}$ around the transit.
\end{itemize}

We refer to the ratio of the transit depth and the standard deviation of fluxes as the signal-to-noise ratio of the transit, $S_{\rm tr} = \delta / \sigma_{\mathrm{flux}}$. The transit depth $\delta$ is calculated based on the latest planet--star radius ratio from scientific literature.

While $432$ planets are known to transit a K- or an M-dwarf, only $99$ of them passed all our selection criteria.
Out of these, $10$~planets were observed with Kepler, $5$~with K2, and $85$~with TESS. One planet (HAT-P-11~b) was observed both with Kepler and TESS with a sufficiently high $S_{\rm tr}$. As some planets in the sample are in the same planetary system, the final sample contains 89 stars, of which 61 are K-dwarfs and 28 are M-dwarfs.

The selected planets had orbital periods ($P_{\mathrm{orb}}$) ranging from $0.8$~d to $482$~d, and a planet--star radius ratio between $0.018$ and $0.304$. The host stars had effective temperatures between $2988$~K and $5358$~K.
The smallest star in our sample had a radius of $R_{*} = 0.21~R_{\odot}$.
The sample stars had Gaia magnitudes within the range $[5.24, 15.71]$.
The planets analysed in this work, along with the number of analysed transits $n_{\mathrm{transit}}$ from each mission, number of Kepler Quarters, K2 Campaigns, or TESS Sectors in which they were observed, as well as parameters adopted from literature, are listed in Table~\ref{tab:analysed_sample} of the Appendix.

\subsection{Transit light curves}

For each planet in our sample, we accessed the  Mikulski Archive for Space Telescopes (MAST) and extracted the light curves of every transit recorded by Kepler or TESS with a cadence of $60-200$~s and $S_{\rm tr} > 2$.
We obtained every observation with a cadence of 60~s from Kepler, and 120~s and 200~s with TESS, using the Lightkurve python package \citep{lightkurve2018}. For Kepler, we accessed the Pre-search Data Conditioning (PDC) light curves, while for TESS we used observations processed with the Pre-search Data Conditioning SAP (PDCSAP) pipeline. Although some data existed with a cadence below 60~s, we did not include such observations because it would have led to increased computational costs without significant improvements in the results due to higher noise levels.

We obtained the times of conjunction $T_0$, orbital period $P_{\rm orb}$, and transit duration $T_{14}$ of each planet based on the latest published values from the NASA Exoplanet Archive. If no transit duration estimate was available, we approximated how long the planet would transit based on how long it would take for the planet to move a distance that corresponds to the diameter of its star. For this, we assumed an inclination of $i=90^\circ$ and zero orbital eccentricity.
We calculated the mid-times $t_c$ of every transit in the Kepler and TESS observations. The mid-time of the $n$th transit occurs at $t_c = T_0 + n \, P_{\rm orb}$. We refer to this number $n$ as the transit index. With these values we then extracted the transit light curves by choosing subsets around the transit mid-times such that data taken at $t_{i} \in [t_c - 1.5\,T_{14}, t_c + 1.5\,T_{14}]$ was selected.
Longer trends, mostly from the rotational modulation of the star were removed by fitting a 2nd order polynomial to the out-of-transit fluxes, and removing it from the whole subset. The fluxes were then divided by the mean flux from the fitted polynomial to maintain uniform transit depths.

The transits of some planets did not occur at the times predicted with the latest $T_{0}$ and $P_{\rm orb}$ estimates. For these objects, we performed a Box-fitting Least Squares \citep[BLS,][]{kovacs2002} analysis in a narrow orbital period range $P_{\mathrm{orb}} \in [P_{\mathrm{lit}}-0.1~\mathrm{d}, P_{\mathrm{lit}}+0.1~\mathrm{d}]$ where $P_{\mathrm{lit}}$ is the last published orbital period (see references in Table~\ref{tab:analysed_sample} of the Appendix). The reason for differing transit times is that additional data has become available for some published transiting planets and this has resulted in small changes in their respective parameters. We have listed the updated orbital periods and times of conjunction in Table~\ref{tab:ephemeris}.

\begin{table}
\caption{Orbital periods $P$ and times of conjunction $T_0$ of planets, obtained through BLS analysis.}              
\label{tab:ephemeris}      
\centering                                      
\begin{tabular}{l r r}          
\hline\hline                        
\multicolumn{1}{c}{Planet} & \multicolumn{1}{c}{$P$ (d)} & \multicolumn{1}{c}{$T_0$ (BJD-2450000)}\\    
\hline                                   
HD 219134 b  & 3.092938     & 8765.954475 \\
HD 219134 c	 & 6.765042     & 8766.169256 \\
HIP 65 A b	 & 0.980972     & 8354.551353 \\
NGTS-1 b	 & 2.646187     & 8469.851517 \\
TOI-1130 b	 & 4.065275     & 9037.939923 \\
TOI-1130 c	 & 8.349502     & 9042.000486 \\
TOI-1278 b	 & 14.475069    & 8711.959279 \\
TOI-1728 b	 & 3.491401     & 8843.275725 \\
TOI-1899 b	 & 29.090104    & 8711.967964 \\
TOI-2202 b	 & 11.914018    & 9089.552796 \\
TOI-674 b	 & 1.977164     & 8544.525107 \\
\hline                                             
\end{tabular}
\end{table}

\section{Statistical analysis}

\subsection{Models}

To detect spot eclipses during the planetary transits, we analysed each transit light curve with two models. The reference model is the analytic transit model of \citet{mandel2002}, with quadratic limb darkening, which corresponds to a case where spot eclipses do not occur during transit.

In addition to the planet's transit, a model with one spot includes a term to describe the change in flux as the planet passes in front of the spot. We empirically model the flux modulation from a spot eclipse as an increase in flux by an amplitude $A$, in the interval $[t_{\mathrm{in}}, t_{\mathrm{out}}]$, with symmetric Gaussian decreases, of variance $\sigma_{\mathrm{spot}}^2$ outside that interval.
At time $t_i$, the flux is modulated by the planet's passage in front of the spot as
\begin{eqnarray}\label{eq:spot_occultation}
  g_{\rm spot}(t_{i}) = \left\{ \begin{array}{ccl}
    A \exp \left [ -\frac{(t_{i} - t_{\rm in})^{2}}{2\sigma^{2}_{\rm spot}} \right ] & \mathrm{ if } & t_{i} < t_{\rm in} \\
    A & \mathrm{ if } & t_{\rm in} \leq t_{i} \leq t_{\rm out} \\
    A \exp \left [ -\frac{(t_{i} - t_{\rm out})^{2}}{2\sigma^{2}_{\rm spot}} \right ] & \mathrm{ if } & t_{i} > t_{\rm out}\,. \\
  \end{array} \right.
\end{eqnarray}
When the planet is not transiting the star, $g_{\rm spot}(t_{i}) = 0$.
A model with $n_{\mathrm{spot}}$ spots is
\begin{equation}
\label{eq:nspot}
    m_{i} = f_{\rm tr}(t_{i}) + \sum_{j=1}^{n_{\mathrm{spot}}} \, g_{{\rm spot},j}(t_{i}) + \epsilon_{i}\,,
\end{equation}
where $f_{\rm tr}(t_{i})$ is the flux from the reference model. The models include a Gaussian noise term $\epsilon_{i}$, with zero mean and a variance of $\sigma_{i}^{2} + \sigma_{w}^{2}$, to account for the instrumental ($\sigma_{i}$) noise and the excess white noise ($\sigma_{w}$) in the observations. As some stars have been observed over an extended period, each of the Kepler quarters and TESS sectors was modelled with a different instrumental noise parameter $\sigma_i$.

To minimise computational cost, we fixed the orbital period, semi-major axis, and stellar radius to the last published values presented on the NASA Exoplanet Archive. We also assumed all orbits to be circular. This assumption has minimal effects on the results, because transit data provides rather weak eccentricity constraints \citep{vaneylen2015}.

We divided the transit light curves of each planet into sets of up to $N$ light curves we call data segments. We modelled these light curves within a subset simultaneously in order to assure the consistency of the free parameters describing the star, the planet, and its orbit. The free parameters are the planet--star radius ratio $r$, inclination $i$, and quadratic limb darkening parameters $u_1, u_2$. The division of transit light curves was designed such that each data segment contained a similar number of subsets, with a maximum of $N = 10$ light curves per segment.

The spot parameters ($A, t_{\mathrm{in}}, t_{\mathrm{out}}, \sigma_{\mathrm{spot}}$) were allowed to change between transits, to account for stellar rotation and spot evolution. The transit mid-time ($t_c$) was also a free parameter, to correct for possible transit-timing variations. We also note that in case a model with $n_{\mathrm{spot}} = n+1$ was found to be considerably better than a model with $n_{\mathrm{spot}} = n$, a model with $n_{\rm spot} = n + 2$ spots was also tested. 

\subsection{Posterior sampling}

To estimate the model parameters and to evaluate the significance of possible detections of spot occultations, we sampled the posterior densities of the parameters, using the Adaptive Metropolis algorithm of \citet{haario2001}. This algorithm is a modified version of the Metropolis-Hastings algorithm \citep{metropolis1953, hastings1970}, that updates the proposal density after each iteration according to the information from the covariance matrix of the vectors generated during the sampling. In practice, this is an approximation of the posterior with a multivariate Gaussian density, which allows for a rapid convergence around the posterior.

We used uniform, uninformative priors for each parameter $\theta$ such that we set them to unity in some interval $\theta \in [a,b]$ and zero outside that interval. When choosing the range of priors for spot parameters, we require that the amplitude of the spot eclipse is smaller than the depth of the transit, to retain the physical nature of the spots and to avoid detecting features caused by some other astrophysical phenomena. The longest time for the spot eclipse signal's maximum amplitude phase (between $t_{\rm in}$ and $t_{\rm out}$) was chosen to be half of the transit duration.
We chose initial values for the sampling randomly from the proposal distribution, save for the inclination, which was set to the latest value in the literature. These inclination values were chosen to prevent scenarios, where the initial value for inclination would be so low that no planetary transit would occur. If no inclination estimate was available for the planet, we used $i=90^\circ$ as an initial value.
Table~\ref{tab:params_sampling} presents the rate at which parameters were allowed to vary, marked as sampling rate (per segments containing $N$ subsets for the parameters of the reference model, per observing run for the noise parameters, per transit for the transit mid-time and the spot parameters) and the intervals of the priors.

\begin{table}[]
\caption{Sampling rates and limits for uniform priors $a,b$, used for the posterior sampling. The index ''lit'' refers to the last published parameter value from the literature.}
\label{tab:params_sampling}
\begin{tabular}{@{}llll@{}}
\hline
\hline
Parameter                & Sampling rate     & $a$                              & $b$ \\ \hline
$r$                      & segment       & $0.7\,r_{\mathrm{lit}}$          & $1.3\,r_{\mathrm{lit}}$ \\
$i~(^\circ)$             &                   & $75$                             & $90$ \\
$u_1$                    &                   & $\mathrm{max}(0, -2 u_2)$        & $1 - u_2$ \\
$u_2$                    &                   & $-1/2 \, u_1$                    & $1-u_1$ \\ \hline
$\sigma_{\mathrm{G}}$   & observing run & $0$                              & $1$ \\ \hline
$t_c$                    & transit       & $-T_{14}/10$                     & $T_{14}/10$ \\
$A$                      &                   & $0$                              & $\delta$                                                  \\
$t_{\mathrm{in}}$        &                   & $t_c-T_{14}/2$                   & $t_{\mathrm{out}}-120~\mathrm{s}$ \\
$t_{\mathrm{out}}$       &                   & $t_{\mathrm{in}}+120~\mathrm{s}$ & $\mathrm{min}(t_{\mathrm{in}}+T_{\rm in}/2$, \\
& & & $t_c+T_{14}/2)$ \\
$\sigma_{\mathrm{spot}}$ &                   & $0$                              & $T_{14}/10$ \\ \hline
\end{tabular}
\end{table}
We sampled each subset with two to four chains, and required that the Gelman--Rubin statistic of each parameter was below $1.05$ to claim that there was no evidence for the non-convergence of the chains and that the identified solutions were unique. A low number of chains was found to be sufficient because most spot eclipses leave an unambiguous signal in the time-series with well constrained solutions in the parameter space. As an example, the corner plots of the sampled parameters of TOI-1268~b, a planet showing one spot occultation event, are shown in Figure~\ref{fig:cornerplot} of the Appendix.

\subsection{Detection thresholds}

To assess the significance of spot eclipses, we compared the models by their likelihood ratios and Bayes factors. We calculated the Bayes factors based on Bayesian information criterion (BIC) estimates \citep{liddle2007}. This statistic has been shown to produce robust results in related statistical problems in astrophysics \citep{feng2016,tuomi2024}.

We interpret the Bayes factors using Jeffreys's scale of probability \citep{jeffreys1961,kass1995}. A model with $n+1$ spots has a strong evidence over a model with $n$ spots if the Bayes factor of the models is over $20$, and decisive evidence, if the Bayes factor is over $150$. This interpretation has been shown to yield reliable spot numbers for a large sample of spot eclipses in the transit light curves of eclipsing binaries \citep{olah2024}. In essence, this approach gains its robustness from the resilience against noisy data. If strong evidence was found in favour of the $n=1$ over the $n= 0$ model for a particular transit, we extended the analysis to a $n = 2$ model, to ensure that no further spots remain undetected.

We considered a signal to be significantly detected with a false alarm probability of 1\% (0.1\%) if the inclusion of the spot occultation event in the model increased the logarithm of the maximum likelihood value by 13.28 (18.47). We adopt this from Wilks's theorem \citep{wilks1938}, which states that the statistic $-2 \log l$, where $l$ is the likelihood ratio, asymptotically follows a $\chi^2$ distribution. All confidence level tests based on $\chi^2$ statistics can be derived from Bayes' rule and Bayesian evidence ratio tests by assuming Gaussian likelihood functions and uniform prior distributions \citep[e.g.,][]{tuomi2013}.

In our application, the Bayes factors based on BIC values are more conservative but the BIC approximation of the Bayes factor can also be sensitive to sudden, local changes in the likelihood function. The posterior sampling may identify vectors in the parameter space of outstandingly high likelihood, that fall relatively far from the mean of the distribution. We visually inspected every detected spot to avoid reporting detections with no perceivable counterparts, caused the most likely by some other astrophysical phenomena or instrumental effects.

\subsection{Spot properties}

The parameters of the spot occultation model can be used to obtain physical parameters for spot groups, such as contrast and radius. Following \citet{morris2017}, the contrast of the dark region can be defined as the ratio between the intensities of the unspotted photosphere ($I_{\rm p}$) and the transited dark region ($I_{\rm sp}$)
\begin{equation}
    c = 1 - \frac{I_{\rm sp}}{I_{\rm p}}\,.
\end{equation}
For dark regions with larger angular radii than the planet, the contrast can be estimated as $c = A/\delta_{\rm p}$, where $A$ is the amplitude of the spot eclipse (Eq.~\ref{eq:spot_occultation}), and $\delta_{\rm p}$ is the depth of the transit at the location of the spot. For dark regions with smaller angular size than that of the planet, the contrast is
\begin{equation}
\label{eq:contrast_correction}
    c = \frac{A}{\delta_{\rm p}} \left ( \frac{R_{\rm p}}{R_{\rm sp}} \right )^{2}\,,
\end{equation}
where $R_{\rm p}$ and $R_{\rm sp}$ are the angular radii of the planet and the spot respectively.

Knowing the effective temperature of the star, the weighted mean effective temperature of the transited dark region can be estimated with the black body approximation \citep{silva2010}
\begin{equation}\label{eq:spot_temp}
T_{\rm s} = \frac{h \nu}{\kappa} \left \{ \log \left [ 1 + \frac{\exp \left ( \frac{h \nu}{\kappa T_{\rm eff}} \right ) - 1}{f} \right ]\right \}^{-1}\,,
\end{equation}
where $\kappa$ and $h$ are the Boltzmann and Planck constants, respectively, $\nu$ is the frequency of the central wavelength of the observing instrument\footnote{The central wavelength of Kepler and TESS telescopes are 640~nm and 786.5~nm, respectively.}, $f = 1-c$ is the spot intensity with respect to the central stellar intensity, and $T_{\rm eff}$ is the stellar effective temperature.

Assuming a stellar inclination of $i = 90^\circ$ and a spin--orbit alignment between the star and the planet, the stellar latitude transited by the centre of the planet is \citep{silva2010}
\begin{equation}
\alpha = \sin^{-1} \left ( \frac{a \cos i}{R_*} \right )\,,
\end{equation}
where $a$ and $i$ are the semi-major axis and the inclination of the planet's orbit, and $R_*$ is the stellar radius. The occulted stellar longitude at some time $t$ is 
\begin{equation}
\theta = \sin^{-1} \left ( \sin \beta \frac{a}{R_*} \cos \alpha \right )\,,
\end{equation}
where $\beta = 2 \pi (t-t_c)/P$.

We chose times $t_{\mathrm{in}}-\sigma_{\mathrm{spot}}$ and $t_{\mathrm{out}}+\sigma_{\mathrm{spot}}$ as the start and end times of the spot eclipse, as we found that they give more realistic estimates for the spot radius than the edges of the constant flux phase, $t_{\mathrm{in}}$, and $t_{\mathrm{out}}$. Accordingly, we define the length of the spot eclipse as $t_{\rm len} = t_{\rm out} - t_{\rm in} + 2 \sigma_{\rm spot}$. Since we cannot determine how close the planet passed to the centre of the dark region, we can only provide a minimum radius $r_{\rm min}$, which corresponds to half the angular width of the eclipsed area.

To test the correspondence between the spot properties obtained from eclipse mapping and the physical spot properties, we modelled the transit light curves of Jupiter in front of the Sun at its activity maximum, as it would be observed from outside the Solar System. These light curves were then sampled with the same methods that we use for sampling the transit light curves of planets in our sample.The comparison is shown in Fig.~\ref{fig:jupiter_transit} and described in detail in Appendix~\ref{appendix:jupiter_transit}. A similar test, though with different orbital parameters has been conducted by \citet{silva2008}.

\begin{figure}
  \resizebox{\hsize}{!}{\includegraphics{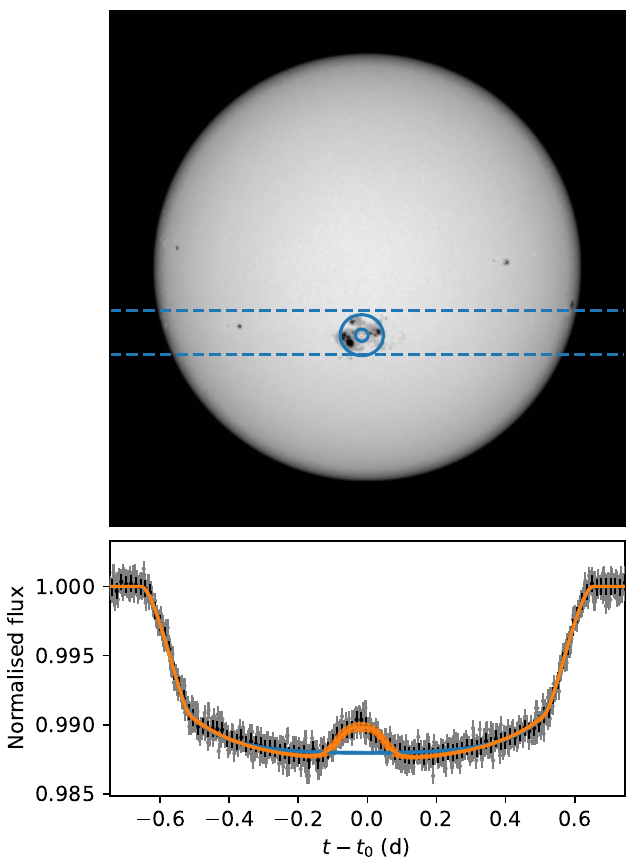}}
  \caption{\textit{Above}:~The Sun during its activity maximum in 2014. The transit path of Jupiter in front of solar latitude $-17.5^\circ$ is marked with dashed lines. The largest sunspot group of the solar cycle, AR2191 is close to the centre of the transit path. The properties of the spot group were reconstructed based on the simulated transit light curve of Jupiter. The inner blue circle indicates the position and size of the spot, if the beginning and end of the spot eclipse are taken as $t_{\mathrm{in}}$ and $t_{\mathrm{out}}$, respectively. The outer circle represents the spot if the eclipse times are adjusted to $t_{\mathrm{in}} - \sigma_{\mathrm{spot}}$ and $t_{\mathrm{out}} + \sigma_{\mathrm{spot}}$. \textit{Below}:~Simulated transit light curve of Jupiter, in front of $-17.5^\circ$ solar latitude. The light curve has similar noise properties as high precision TESS observations. A 1- (orange) and 0-spot (blue) model is shown on top of the light curve.}
  \label{fig:jupiter_transit}
\end{figure}

This simulation shows that the spot temperature and coverage estimated using transit mapping, may differ considerably from reality. Thus, when we refer to spot temperatures  in this study, it is actually a weighted mean temperature representing the whole active region covered by the transit. This also refers to the spot coverage: rather than representing the total umbral area, it refers to the the whole active region, for which the temperature is estimated.

\subsection{False positives \label{sec:false_positives}}

The posterior samplings and model comparisons probably yield robust results. However, they are not infallible. If in some cases, complex correlated variation is present in the observations, it may get mistaken for a spot eclipse signal. Therefore, it is necessary to have some safeguards to ensure that the identified signal comes from a spot eclipse. Astrophysical and instrumental false positives can also mimic signals from spot occultations. For this reason, we tested whether the flux variations, identified as spot eclipses have other origins, by performing a series of tests for false positives.

Although small-scale brightness variations on the stellar surface, or simply correlated noise, can give rise to false positives, the brightness modulations identified as spot eclipses, can also be caused by some unknown, unrelated astrophysical source. Yet, in such a case, similar modulations in flux would also be present in the time-series outside the transit events. We tested this possibility for the stars that showed strong evidence for spot occultations, by extracting $N=50$ randomly selected subsets from their light curves when no planetary transit occur and by searching those out-of-transit subsets for signals described in Eq.~\ref{eq:spot_occultation}.

Flares may cause variations similar in amplitude to spot eclipses. Distinguishing between flares and spot eclipses is a relatively straightforward task, as flares follow distinct asymmetric morphologies \citep{howard2022}, whereas spot eclipses are expected to be more symmetric. We tested the possibility that the brightness variations identified as spot eclipses may be due to flares by sampling a simple flare model on the transit light curves with spot detections. Our flare model consisted of a discontinuity at time $t_{\rm f}$, such that the flux jumps above the baseline, followed by an exponential decay.
At time $t_i$ the flux is modulated by the flare as
\begin{equation}
g_{\rm flare}(t_{i}) =
\left\{ \begin{array}{ccl}
    0 & \mathrm{ if } & t_{i} < t_{\rm f} \\
    A_{\rm f} \exp \left ( - \frac{t_i - t_{\rm f} }{\tau} \right ) & \mathrm{ if } & t_{i} \geq t_{\rm f}
  \end{array} \right.
\end{equation}
where $A_{\rm f}$ is the amplitude of the flare at peak time $t_{\rm f}$, and $\tau$ is the timescale of the exponential decay. A planetary transit model with a co-occurring flare is then of shape
\begin{equation}
    m_i = f_{\rm tr} (t_i) + g_{\rm flare}(t_{i}) + \epsilon_i\,,
\end{equation}
where $f_{\rm tr}$ is the flux from the reference model and $\epsilon_i$ is the Gaussian noise term (see Eq.~\ref{eq:nspot}). We rejected spot eclipse candidates after visual inspection and if the Bayes factor of the flare model in favour of the spot model indicated a strong evidence of a flare signal.

The pre-search data conditioning pipeline of the Kepler and TESS telescopes may introduce artefacts to the observations on similar timescales as spot eclipses. To ensure that the identified spot eclipse signals are not artefacts from the pipeline, we sampled the SAP light curves for spot eclipses, where such events were found.

Light curves on longer timescales also provide some additional means for testing the nature of the signals interpreted as spot occultation events. Assuming that the spots identified from eclipses are non-axisymmetric with respect to the stellar axis of spin, we expect that they leave a photometric signal in the light curve of the star as they co-rotate with the stellar surface. The maximum flux deficit a dark region with contrast $c$ and radius $r_{\rm spot}$ can leave in the light curve of the star is \citep{tregloan2019}
\begin{equation}
\Delta F_{\rm spot} = c \, (1 - \cos r_{\rm spot})\,.
\end{equation}
This approach gives an upper limit for photometric variations on the timescale of the stellar rotation, induced by the co-rotating spot. If the expected signal is above the noise limit for rotational modulation, but cannot be identified in longer baseline observations, the in-transit flux variation may have some other origin than a spot eclipse. 

\subsection{Average spot number and spot filling factor}

The average number of dark active regions on a star can be approximated, assuming a uniform spot distribution, by taking the occurrence rate of spot eclipses and the size of the segment on the star that is transited by the planet. This is a very crude approximation, which assumes a uniform latitude distribution of dark regions, and it does not, for example, accurately reflect the Sun's spot distribution.

Let us take the radius of a star to be unity and let us denote the average number of dark active regions on the stellar surface as $N_{\mathrm{s}}$. At any time, we see half of the stellar surface, so the average number of spots on the visible hemisphere is $1/2 N_{\mathrm{s}}$.

As the planet transits the stellar hemisphere, the area of the transited band on the stellar surface, without stellar rotation, is
\begin{equation}
    A_{\rm tr} = \pi h\,,
\end{equation}
where $h = \mathrm{min}(b+r,1) - \mathrm{max} (b-r,-1)$, and $b = \sin \alpha$ is the impact parameter. The number of transited dark regions is obtained by dividing the transited area by the surface area of the visible hemisphere, and multiplying it by the number of dark regions on the visible hemisphere
\begin{equation}
    N_{\rm{s,tr}} = \frac{N_{\rm s}}{2} \frac{\mathrm{min}(b+r,1)-\mathrm{max}(b-r,-1)}{2}\,.
\end{equation}

\begin{figure}
    \centering
    \includegraphics[width=.5\textwidth]{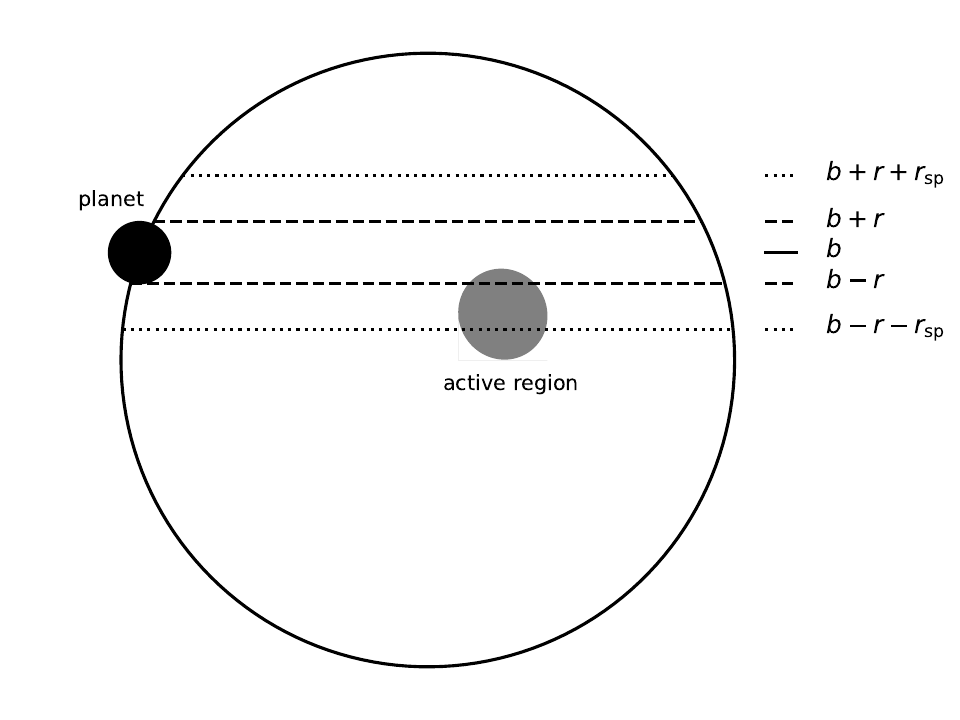}
      \caption{
      The geometry of a planetary transit across a star. The planet of radius $r$ transits the star at impact parameter $b$. The band transited by the planet is marked by dashed lines. Active regions with radii $r_{\rm sp}$ and centres located within the band outlined by dotted lines may be eclipsed by the planet.
              }
         \label{fig:eclipse_band}
   \end{figure}

However, assuming that the centres of some active regions are not transited by the planet (grazing transits), the size of the band where spots may be eclipsed widens on both sides by the radius of the active region $r_{\rm sp}$, and the number of eclipsed dark regions is described by
\begin{equation}
    N_{\rm s,tr} = \frac{N_{\rm s}}{2} \frac{\mathrm{min}(b+r+r_{\rm sp},1) - \mathrm{max}(b-r-r_{\rm sp}, -1)}{2}
    \,.
\end{equation}
This is a valid consideration, as we find that the apparent radii of the eclipsed dark regions are often larger than the apparent radius of the planet.
It follows from this calculation, that the average spot number is not necessarily an integer. Fig.~\ref{fig:eclipse_band} shows the geometry of the transit.
In summary, the mean number of spots for the star is calculated by dividing the number of detected spot eclipse candidates by the number of analysed transits, and multiplying that result with the inverse ratio of the area of the band from where spots may be transited by the planet to the area of the full stellar surface.

The spot filling factor of the star, $f_S$, can be approximated by multiplying the average number of dark regions with the mean area of dark regions and dividing it with the area of the stellar surface. This approximation is valid even if there is no spin-orbit alignment, and the inclination of the star is not $90^\circ$.

\section{Results}

After analysing $3273$ light curves containing transit events of $99$ planets from the Kepler, K2, and TESS missions, we found six targets that show strong evidence for eclipse events over dark regions during the transits. In total, we identified $105$ such candidate eclipses.

\begin{figure*}
   \centering
   \includegraphics{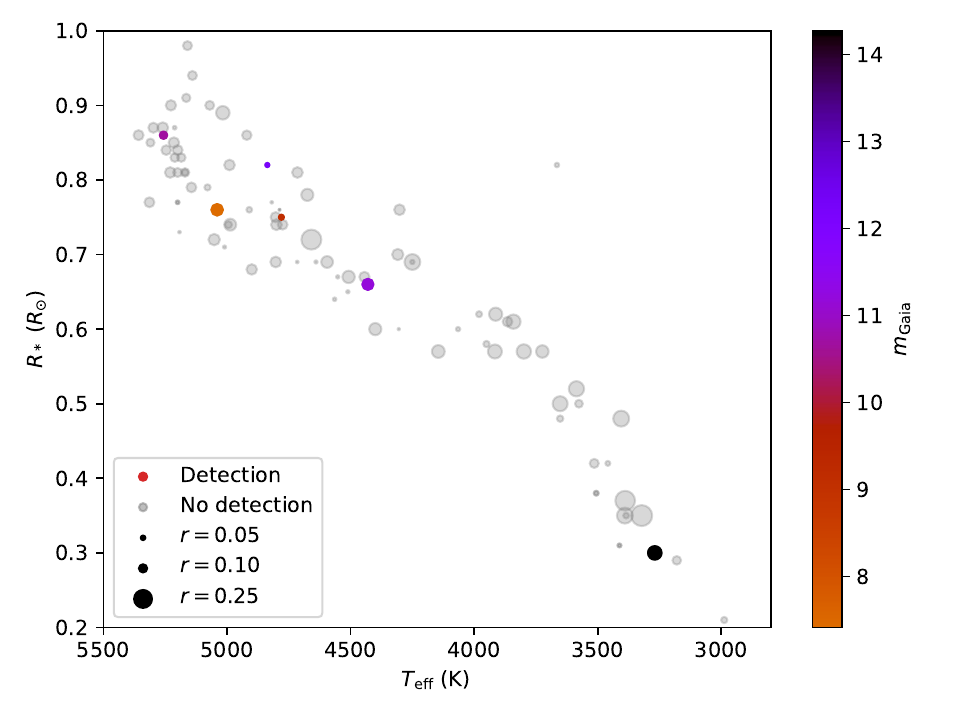}
   \caption{Distribution of systems with spot eclipse detections (coloured dots) in terms of stellar effective temperature $T_{\mathrm{eff}}$, stellar radius $R_*$, and planet--star radius ratio $r$, compared to systems where no spot eclipse was found (grey dots). The colour mapping of the systems with spot eclipses corresponds to the Gaia magnitudes $m_{\mathrm{Gaia}}$ of the stars.
         \label{fig:distribution}}
    \end{figure*}

In Fig.~\ref{fig:distribution}, we show the sample of analysed exoplanets on the stellar effective temperature--stellar radius plane. The scaling of the filled circles corresponds to the planet--star radius ratios. Targets for which spot eclipses were identified are marked in colours. The colouring corresponds to the Gaia magnitudes $m_{\rm Gaia}$. As stellar brightness correlates with the photon counts and the noise levels of the observations, brighter stars are expected to show more spot occultations than fainter ones. Grey dots mark the planets where no spot eclipses were identified. All but one star with a spot eclipse candidate was of K spectral class.

In Table~\ref{tab:spot_properties}, we provide detailed information about each planetary transit (except for those of HAT-P-11~b, see Section~\ref{sec:hatp11b}) during which a candidate spot eclipse occurred. We list the planet's name, the index and the mid-time ($t_c$) of the transit, the significance of the spot eclipses expressed as the logarithm of the likelihood ratio ($\log l$) between the spot model and the reference model, and the Bayes factors between the two models ($B_{1,0}$). Additionally, four parameters characterising the spot eclipses are reported: the length of the spot eclipse ($t_{\rm len} = t_{\rm out}-t_{\rm in} + 2 \sigma_{\rm spot}$), the mid-time of the spot eclipse relative to the time of conjunction of the planetary eclipse ($t_{\rm mid}$), the minimum angular size of the eclipsed region ($r_{\rm min}$), and the contrast of the eclipsed region ($c$).

\begin{table*}
\caption{Significance and properties of candidate spot eclipse events identified in this analysis. \label{tab:spot_properties}}
\begin{center}
\begin{tabular}{lrrrrrrrr}
\hline \hline 
\multicolumn{1}{c}{Planet} & \multicolumn{1}{c}{Transit index} & \multicolumn{1}{c}{\begin{tabular}[c]{@{}c@{}}$t_c$\\ (BJD-2450000)\end{tabular}} & \multicolumn{1}{c}{$\log l$} & \multicolumn{1}{c}{$\log_{10} B_{1,0}$} & \multicolumn{1}{c}{\begin{tabular}[c]{@{}c@{}}$t_{\rm len}$\\ (s)\end{tabular}} & \multicolumn{1}{c}{\begin{tabular}[c]{@{}c@{}}$t_{\rm mid}$\\ (s)\end{tabular}} & \multicolumn{1}{c}{\begin{tabular}[c]{@{}c@{}}$r_{\mathrm{min}}$\\ ($^\circ$)\end{tabular}} & \multicolumn{1}{c}{$c$} \\
\hline
HD 189733 b & $2161$ & 9426.59 & 17.96 & 3.37 & $932_{-199}^{+213}$ & $-509_{-123}^{+87}$ & $4.5_{-1.0}^{+1.1}$ & $0.038_{-0.016}^{+0.029}$ \\  
HD 189733 b & $2165$ & 9435.46 & 18.41 & 3.55 & $948_{-146}^{+196}$ & $1890_{-82}^{+65}$ & $5.0_{-0.8}^{+1.0}$ & $0.076_{-0.025}^{+0.041}$ \\ 
HD 189733 b & $2167$ & 9439.90 & 35.01 & 10.77 & $1989_{-300}^{+300}$ & $-1254_{-119}^{+128}$ & $10.1_{-1.6}^{+1.6}$ & $0.039_{-0.005}^{+0.004}$ \\ 
HD 189733 b & $2168$ & 9442.12 & 41.72 & 13.68 & $3164_{-245}^{+237}$ & $690_{-128}^{+112}$ & $15.7_{-1.3}^{+1.3}$ & $0.031_{-0.003}^{+0.003}$ \\ 
HD 189733 b & $2170$ & 9446.55 & 36.83 & 11.68 & $1257_{-211}^{+173}$ & $414_{-58}^{+65}$ & $6.0_{-1.0}^{+0.8}$ & $0.058_{-0.007}^{+0.011}$ \\ 
HD 189733 b & $2317$ & 9772.69 & 19.15 & 3.87 & $1598_{-254}^{+164}$ & $-496_{-128}^{+85}$ & $7.7_{-1.2}^{+0.8}$ & $0.027_{-0.006}^{+0.005}$ \\ 
HD 189733 b & $2323$ & 9786.00 & 28.46 & 7.92 & $4409_{-152}^{+118}$ & $-169_{-234}^{+249}$ & $21.7_{-0.8}^{+0.7}$ & $0.030_{-0.003}^{+0.003}$ \\ 
HD 189733 b & $2326$ & 9792.65 & 22.66 & 5.52 & $1946_{-1139}^{+1306}$ & $-716_{-1006}^{+1232}$ & $9.8_{-5.6}^{+6.7}$ & $0.006_{-0.004}^{+0.013}$ \\ 
Kepler-411 c & $116$ & 5877.01 & 17.99 & 1.02 & $921_{-728}^{+806}$   & $-290_{-225}^{+163}$  & $2.5_{-1.0}^{+1.0}$ & $0.318_{-0.012}^{+0.013}$ \\ 
Kepler-411 c & $117$ & 5884.85 & 21.66 & 3.91 & $1964_{-1438}^{+1500}$ & $-2655_{-245}^{+315}$ & $6.9_{-1.6}^{+1.5}$ & $0.204_{-0.005}^{+0.004}$ \\ 
Kepler-411 c & $121$ & 5916.18 & 20.06 & 3.26 & $734_{-631}^{+535}$   & $-1903_{-149}^{+168}$ & $2.2_{-0.8}^{+0.8}$ & $0.321_{-0.010}^{+0.010}$ \\ 
TOI-1268 b & $3$ & 8728.06 & 24.05 & 5.34 & $5272_{-544}^{+550}$ & $1561_{-250}^{+252}$ & $24.2_{-2.7}^{+2.9}$ & $0.159_{-0.012}^{+0.021}$ \\ 
TOI-3884 b & $-16$ & 9570.15 & 14.87 & 2.12 & $2179_{-667}^{+363}$ & $-1079_{-175}^{+152}$ & $29.5_{-9.8}^{+6.9}$ & $0.306_{-0.056}^{+0.068}$ \\ 
TOI-3884 b & $-15$ & 9574.69 & 15.23 & 2.27 & $1469_{-328}^{+501}$ & $-917_{-167}^{+193}$ & $18.3_{-4.3}^{+6.5}$ & $0.333_{-0.073}^{+0.083}$ \\ 
TOI-3884 b & $1$   & 9647.41 & 13.92 & 1.70 & $1898_{-449}^{+363}$ & $-938_{-174}^{+147}$ & $24.1_{-5.6}^{+5.1}$ & $0.315_{-0.056}^{+0.063}$ \\ 
WASP-107 b & $4$  & 7607.22 & 160.95 & 64.88 & $937_{-140}^{+74}$ & $-811_{-7}^{+12}$ & $6.2_{-0.6}^{+0.5}$ & $0.211_{-0.005}^{+0.004}$ \\ 
WASP-107 b & $6$  & 7618.66 & 134.47 & 53.01 & $850_{-76}^{+69}$ & $-1272_{-23}^{+85}$ & $5.7_{-0.6}^{+0.5}$ & $0.158_{-0.003}^{+0.003}$ \\ 
WASP-107 b & $11$ & 7647.27 & 28.17 & 6.85 & $2961_{-804}^{+843}$ & $-42_{-325}^{+300}$ & $19.6_{-5.5}^{+6.0}$ & $0.017_{-0.004}^{+0.004}$ \\ 
\hline

\hline \hline
\end{tabular}
\end{center}
\end{table*}

\subsection{Notes on individual objects}

\subsubsection{HAT-P-11 \label{sec:hatp11b}}

HAT-P-11 is a $0.81_{-0.03}^{+0.02}$ solar mass, K4V spectral type star with an effective temperature of $4780 \pm 50$~K. It is transited by one planet, HAT-P-11 b, a super-Neptune on a $4.89$-day orbit \citep{bakos2010}. Spot occultations by the planet have been found in its transit light curves soon after its discovery \citep{sanchis2011}. Another planet in the system was detected from radial velocity measurements on a $\sim 3400$~d orbit \citep{yee2018}. The star was observed by Kepler in its original mission, and then by TESS over six sectors. Its relatively high brightness ($V = 9.46$) makes it an ideal target for monitoring with both instruments.

\citet{bakos2010} found significant flux variations in the observations of HAT-P-11, with a period of $P=29.2$~d, which they attribute to the rotational period of the star. Age estimates for the star range between $2.69_{-1.24}^{+2.88}$~Gyr \citep{morton2016} and $6.5_{-4.1}^{+5.9}$~Gyr \citep{bakos2010}. From measuring the Rossiter-McLaughlin effect, \citet{winn2010} estimated a sky-projected obliquity of $103_{-10}^{+26}$~degrees for the system.

We analysed $214$ transits of HAT-P-11~b, $183$ recorded by Kepler, and $31$ recorded by TESS. We identified 87 possible spot occultation events by HAT-P-11~b, all of them observed by Kepler. This planet alone has more spot eclipse candidates than all other planets in our sample combined, and we list its spot eclipse candidates in Table~\ref{tab:hatp11bspots} of the Appendix. The planet was also the only one in our sample that showed multiple spot eclipses during a single transit.

We found extreme variations around some transit events in the PDC light curves as artefacts from the Kepler Pre-search Data conditioning Pipeline. These variations complicated the normalisation of certain transit light curves, forcing us to exclude some light curves from our sample. Accounting for these variations is outside the scope of the current work. For this reason, we advise caution when interpreting the statistics of HAT-P-11~b, and suggest that the number of identified spots provides a lower limit to the number of identifiable spot occultation events in the sample.

The contrasts of the eclipsed active regions were between $0.040_{-0.005}^{+0.005}$ and $0.688_{-0.021}^{+0.053}$. Adopting the effective temperature estimate of $4780 \pm 50$~K \citep{bakos2010}, we calculate that the mean temperature of the darkest transited region was $\Delta T \approx 950$~K cooler than the unspotted photosphere.

\subsubsection{HD 189733}

HD~189733 is the brightest star ($m_{\mathrm{Gaia}} = 7.41$) for which we detected spot eclipse candidates. The star is of spectral type K2V 
\citep{gray2003} and has an effective temperature of $T_{\mathrm{eff}} = 5050 \pm 50$~K \citep{bouchy2005}. It is the main component of a binary system, where the projected separation of the two components is about 216 AU \citep{bakos2006}. The star is known to be orbited by one planet, a hot Jupiter with an orbital period of $P = 2.22$~d.
The planetary orbit has an inclination angle $i = 85.5^\circ$ with a low mutual inclination angle 
$\lambda = -0.85^\circ$ \citep{triaud2009}.

\citet{bonomo2017} estimated a stellar rotation period of $11.95 \pm 0.01$~d from ground-based photometric observations.
Age estimates for this star range from less than $1$~Gyr \citep{ghezzi2010} to $6.8^{+5.2}_{-4.4}$~Gyr \citep{torres2008}.
The star is known to show signals of relatively high stellar activity.  \citet{wright2004} measured a chromospheric activity index of $S = 0.525$, while \citet{barnes2016} reported $\log R'_{HK}$ indices between $-4.55$ and $-4.50$. Both statistics imply increased magnetic activity, which would favour a younger age.

The system has exhibited one of the early examples of spot eclipses by an exoplanet. \citet{pont2007} found two such events while monitoring three transits of the planet with the Hubble Space Telescope (HST). Later, \citet{sing2011} found similar spot eclipse signals in repeated HST observations, and, through stellar atmospheric modelling, estimated the spot temperatures to be around $4250$~K.

Recently, upon analysing archival spectra from HST observations, \citet{narrett2024} inferred the spot-covering fraction and the effective temperature of the spots for HD~189733. They found spot covering fractions between $38\% \pm 4\%$ and $47\% \pm 3 \%$ for spectra taken at different times, and a spot effective temperature of $3222^{+100}_{-116}$~K.

\begin{figure}[]
    \centering  \includegraphics[width=.5\textwidth]{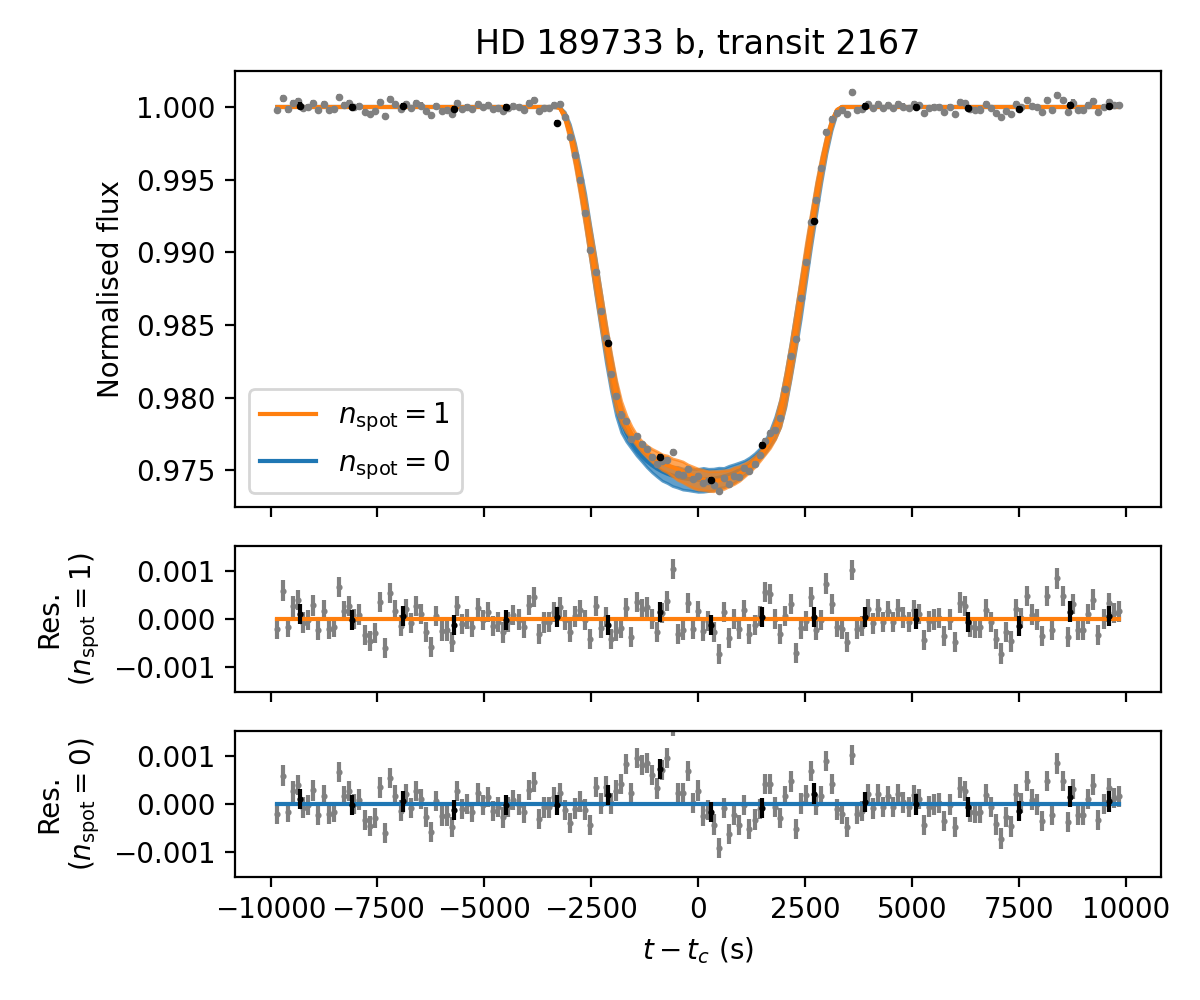}
      \caption{Transit of HD 189733 b observed by TESS with a cadence of 120~s. Black dots represent the mean of consecutive subsets of ten observations. Maximum likelihood fits from the $n_{\rm spot} = 1$ and $n_{\rm spot} = 0$ models are marked with orange and blue lines respectively, one sigma uncertainties are marked as shaded areas. The residuals of the highest likelihood $n_{\rm spot} = 1$ and $n_{\rm spot} = 0$ model fits are shown in the middle and the bottom of the figure, respectively.
      The feature between $t-t_c \approx -2200$~s and $t-t_c \approx -300$~s is an eclipse of a starspot with a radius of $r_{\rm min} = 9.8_{-2.2}^{+4.5} \, ^\circ$ and a contrast of $0.04$.}
         \label{fig:transits_hd189733b1}
   \end{figure}

The target was observed by TESS in Sectors~41 and 54. We analysed the 21 transits recorded by TESS and detected eight new spot eclipses in the transits of HD~189733~b. After comparing the spot signals with the simple flare model, described in Section~\ref{sec:false_positives}, we rejected one detection (transit~2324). Fig.~\ref{fig:transits_hd189733b1} shows an example transit light curve of HD~189733~b with one spot eclipse, shortly before the centre of the planetary transit.

The contrasts of the eclipsed regions we detected are relatively low, ranging from $0.006_{-0.004}^{+0.013}$ to $0.076_{-0.025}^{+0.041}$. The detection of such low contrast eclipses was possible due to the high signal-to-noise ratio ($S_{\rm tr} = 57.3$) of the observations.
The contrast of the darkest eclipsed region corresponds to an effective temperature difference of $\Delta T \approx 120$~K, when taking $5050 \pm 50$~K as the effective temperature of the star \citep{bouchy2005}. This, however, does not strictly correspond to the temperature difference of the photosphere and the spot umbra itself, as the temperature difference estimate is based on the weighted mean effective temperature of the entire eclipsed region, which may include unspotted areas or even bright faculae.

The durations of spot eclipses were between $\sim 930$~s and $\sim 4410$~s, and the minimum radii of the eclipsed regions ranged from $4.5_{-1.0}^{+1.1}$ degrees to $21.7_{-0.8}^{+0.7}$ degrees. These properties are consistent with earlier HST observations by \citet{pont2007}.
We estimate an average number of dark regions of $N_S = 3.79$, which, with the maximum spot radius of the sample gives a spot filling factor of $13\%$.

When assessing the frequency of astrophysical false positives, we found one out-of-transit light curve with strong evidence for a signal resembling spot eclipse. This is possibly due to the elevated activity of the star, and in our view does not invalidate the spot eclipses found during the planetary transits, due to their vastly differing occurrence frequencies during and outside of transit.

Analysing the transit observations previously published, \citet{baluev2022} reported transit timing variations for HD~189733~b on the order of $\sim 70$~s. Their analysis suggests that only $\sim 10$~s of these variations can be attributed to the planet transiting in front of an uneven photospheric brightness field. We find a maximum transit mid-time difference of $\Delta t_c \sim 30$~s in the TESS observations. The largest discrepancy in transit mid-times between the $n_{\rm sp} = 0$ and $n_{\rm sp} = 1$ models was $\sim 20$~s.

\subsubsection{Kepler-411}

Kepler-411 is a K2V spectral type star of $0.87 \pm 0.04$ solar masses and the age of $212 \pm 31$~Myr \citep{sun2019}. The star is orbited by four planets, of which three transit. The inclination of the star is $89.79\pm0.19 ^\circ$ \citep{tuomi2024}.

The star is well known for its activity and spot eclipses from its planets. \citet{araujo2021} studied the star extensively in terms of spot eclipses. Their detection criteria for spot eclipse was that the amplitude of variations from the reference model exceed $3 \sigma$ and reported nearly two hundred spot eclipses in the Kepler time-series.

Recently, \citet{tuomi2024} studied the star in terms of spot signatures. They sampled the transit light curves of planet~c for spot eclipses and performed rotational modelling to characterise spot geometries for the whole stellar surface. They applied the same detection criteria in their analysis as applied in the current work and reported three spot eclipse events from planet~c, all within a $50$~d long interval.

The star is in the original Kepler field and was observed in several TESS Sectors. The noise levels of the TESS observations were too high to enable robust spot detections. We therefore searched the Kepler data for spot occultation events in the transits of all the planets but managed to detect spot eclipses only for the transits of planet c. This may be explained by the lower number of transits from the other planets, the shallowness of their transits, or the spots being constrained to the latitude band coinciding with the orbital inclination of the planet~c.

After comparing the spot signals to the simple flare model, we rejected one eclipse detection over a dark region (transit 164). Our spot detections correspond to those discussed in \citet{tuomi2024}. We could not replicate most spot detections of \citet{araujo2021}, probably due to adopting a rather conservative set of detection criteria.

\subsubsection{TOI-1268}

TOI-1268 is a young, K1--K2~V spectral type main sequence star, with an estimated age of $245 \pm 135$~Myr \citep{dong2022}. It has a Gaia magnitude of $m_{\rm Gaia} = 10.69$ and a TESS magnitude of $m_{\rm TESS} = 10.15$. The TESS transit observations have a mean $S_{\rm tr}$ of $6.17$. The star's rotation period is $10.8$ days \citep{dong2022}. TOI-1268 is orbited by a known planet (TOI-1268~b) with a mass comparable to Saturn. The planet has an orbital period of $P_{\rm orb} = 8.16$ days \citep{subjak2022}. Measurements of the Rossiter--McLaughlin effect by \citet{dong2022} indicate that the star and its planet are in spin--orbit alignment.

\begin{figure}[]
    \centering
    \includegraphics[width=.5\textwidth]{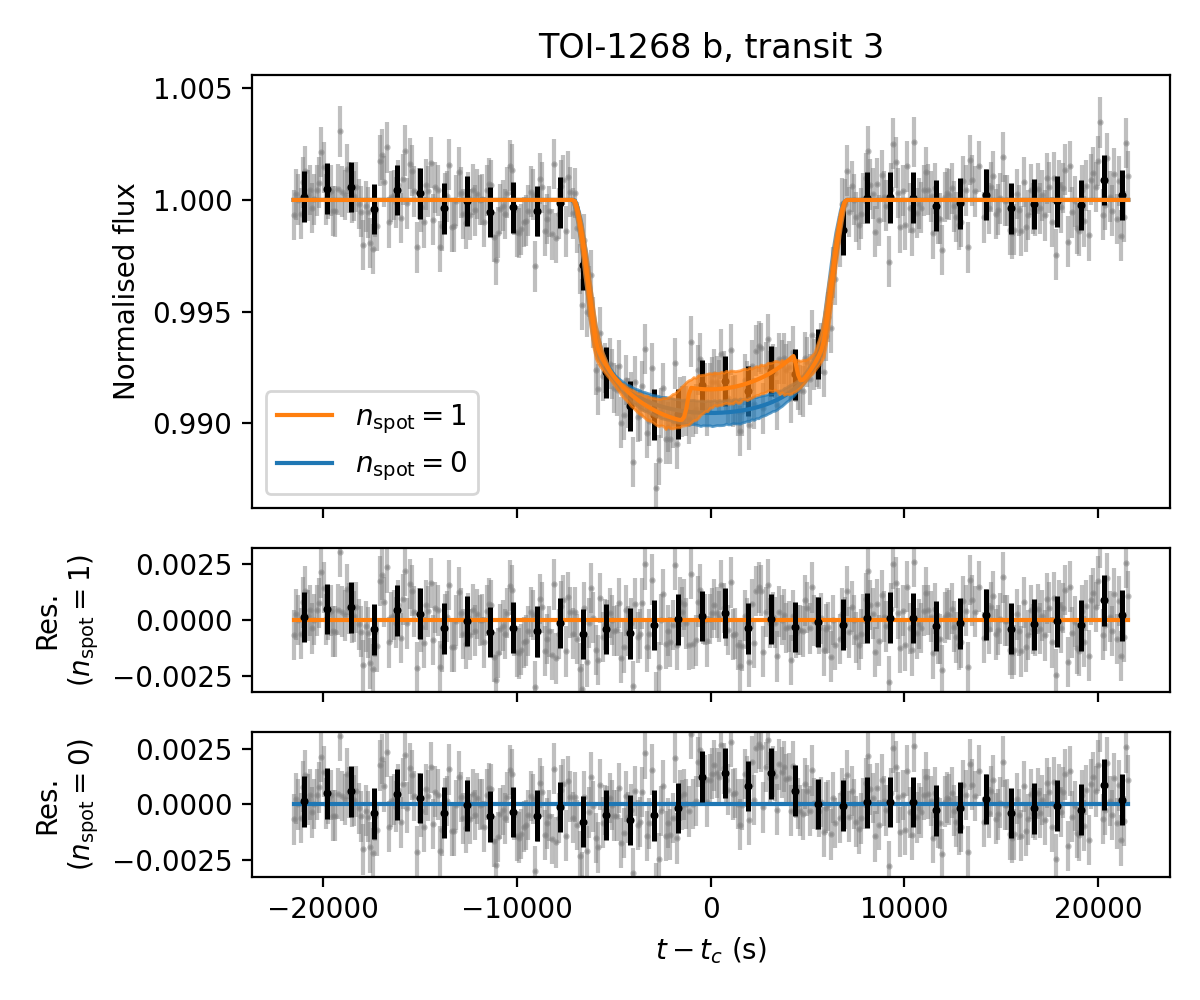}
      \caption{Transit of TOI-1268 b observed by TESS with a cadence of 120~s. The spot eclipse occurs between $t-t_c \approx -1100$~s and $t-t_c \approx 4200$~s. The eclipsed spot region has a minimal radius of $r_{\rm min} = 24.5^{+11.4}_{-5.5}\,^\circ$ and a contrast of $0.16^{+0.02}_{-0.01}$. }
         \label{fig:transits_toi1268b1}
   \end{figure}

We analysed $14$ transits of the planet, observed over five TESS Sectors. We obtained decisive evidence for one spot eclipse ($\log_{10} B = 5.34$), during transit~3 of the planet. The transit light curve with the residuals from the spot- and the reference models is shown on Fig.~\ref{fig:transits_toi1268b1}.

The minimal radius of the eclipsed dark region is $24.2_{-2.7}^{+2.9} \, ^\circ$, while its contrast is $0.159^{+0.021}_{-0.012}$. This radius is towards the higher end of the spot radius distribution reported in this work, and is the largest spot radius for a star with non-axisymmetric signals. Taking the effective temperature of the star to be $5257 \pm 40$~K \citep{dong2022}, we obtain a mean effective temperature of the eclipsed active region as $5031 \pm 40$~K. We note that the Bayesian analysis indicated that the flare model has strong evidence in favour of the spot model ($B = 31.1$), but after the visual inspection of the signal, we find it unlikely that the brightening during the transit was caused by a flare.

Based on visual inspection of the eclipse signal it may appear credible that the signal originates from two separate spots. However, the sampling of a model with $n_{\rm spot} = 2$ did not provide strong evidence for the presence of two distinct spots.

\subsubsection{TOI-3884}

TOI-3884 is an M4 spectral class star, transited by one planet with an orbital period of $P = 4.54$~d, on a supposedly highly misaligned ($\psi = 50 \pm 12 ^{\circ}$) orbit \citep{almenara2022}. Rotation period estimates for TOI-3884 range between $P_{\mathrm{rot}} = 4.22 \pm 1.09$~d \citep{libby2023} and $P_{\mathrm{rot}} \approx 18$~d \citep{almenara2022}. Both estimates rely on the rotational broadening of lines in the star's spectra, since no rotational modulation can be observed in the light curve of the star. The former value would, in practice, suggest a synchronous orbit for the planet. The estimated short rotation period suggests the star is reasonably young.

The star was observed by TESS with a cadence of 120~s in two consecutive sectors, during which eight transits of TOI-3884~b were recorded. Shortly after each transit began, the light curve was modulated with a feature similar to what a spot occultation would cause. Based on TESS measurements, and observations taken with a 1-m telescope at Teide observatory with the Las Cumbres Observatory Global Telescope (LCOGT), \citet{almenara2022} interpreted that the modulation is caused by a giant polar spot of radius $r=49^{\circ}$ (17\% of the surface covered). Subsequent ground-based observations of \citet{libby2023} confirmed the existence of the spot.

We performed our analysis on the eight transits recorded by TESS and found evidence for three spot signatures. We found that the 1-spot model was strongly favoured against the 0-spot model for two transits ($B = 131.3$, $B = 50.1$ for transits $-16$ and 1, respectively). We also found one transit for which the 1-spot model had decisive evidence against the 0-spot model ($B = 185.7$ for transit $-15$).

We note that we obtained solutions that remained reasonably similar between transits ($r_{\textup{min}} = 25.4_{-6.4}^{+4.4}$~deg, $c = 0.31 _{-0.07}^{+0.03}$, on average for the eight transits). This suggests that the varying detection statistics are probably an artefact of the high noise levels of TESS observations.

\subsubsection{WASP-107}

WASP-107 is a K6V spectral type star. It is orbited by two planets, of which one is transiting \citep{anderson2017}. WASP-107~b orbits its star with a $5.72$~d period. Using Rossiter--McLaughlin measurements, \citet{rubenzahl2021} found that WASP-107~b has a retrograde, polar orbit, with a projected spin-orbit angle of $| \lambda | = 118^{+38}_{-19} \, ^\circ$.

WASP-107 has a V-band brightness of $11.47$. It was observed in one K2 campaign with 60~s cadence with a high signal-to-noise ratio ($S/R = 35.7$). Spot eclipses have been reported for this star by several authors with differing criteria for detection. The star has not been observed by TESS with a sufficiently high cadence that would allow the detection of spot eclipses.

\citet{mocnik2017} inspected the residuals of the transit light curves of WASP-107 b, after the best-fitting MCMC transit model had been subtracted. They inspected the residuals by eye and categorised them as definite and possible spot occultation events. They found five definite and four possible spot occultations. \citet{dai2017} found three spot eclipses (which they called ``visually obvious'') in the same K2 observations. Their spot model was of a simple Gaussian curve, and they demanded that the $\Delta {\rm BIC} > 10$ in favour of the spot model, as spot detection criteria. Both \citet{mocnik2017} and \citet{dai2017} note that the star and the planet are not in spin--orbit alignment.

After analysing nine transits recorded by K2, we identified three potential spot eclipses in the transits of WASP-107~b. The spots had contrasts between $0.017_{-0.004}^{+0.004}$ and $0.211_{-0.005}^{+0.004}$, and minimal radii ranging from $5.7_{-0.6}^{+0.5}$ degrees to $19.6_{-5.5}^{+6.0}$. Notably, the first two of the three eclipsed regions were smaller and had higher contrasts compared to the third, implying that the last eclipsed feature is warmer and more diffuse than the first two. The spots identified by us correspond to those identified by \citet{dai2017}.

Adopting a stellar effective temperature of $4430 \pm 120$~K \citep{anderson2017}, the contrasts indicate that the effective temperatures of the eclipsed regions range from $4235 \pm 110$~K to $4413 \pm 119$~K. We estimate the average number of dark regions on the star to be $3.73$ and the upper limit for the spot filling factor to be $11\%$.

\section{Discussion}

The spot eclipse modelling and significance assessment was performed following a Bayesian framework common in related fields of astrophysics. The criteria for confirming a more complex model were more conservative than those used by most authors. For this reason, we detected fewer spots for several stars with published spot detections. This is particularly clear for e.g., Kepler-411 \citep[see also][]{tuomi2024}, but we also failed to detect the reported spot eclipses of \citet{zaleski2020} for Kepler-45~b. This does not imply that the corresponding spot detections reported earlier were false positive detections, but it does cast doubt on their statistical significance.

Detected spot eclipse durations range from $\sim 200$~s to $\sim 5300$~s. As shown by \citet{silva2010}, spots closer to the central meridian are more detectable than those on the stellar limbs. This is due to distortion effects, as the transit light curve becomes steeper as the planet transits in front of regions closer to the stellar disk. We found $90\%$ of spot candidates within $22.4^\circ$ of the meridian, with none beyond $46.9^\circ$.

For planets with smaller angular radii than that of the transited dark regions, the contrast did not exceed $0.43$, while the same number was $0.69$ for smaller features, where the contrast was corrected following Eq.~\ref{eq:contrast_correction}. If the contrast of the dark region is $c \lesssim 0.1$, one-sigma errors for the stellar dark region and unspotted surface temperatures overlap. The dark region's effective temperature is derived from the stellar surface temperature and the contrast of the eclipsed region, with the errors propagating accordingly. The real temperature difference between individual spots and the photosphere may be an order of magnitude higher than inferred from the eclipsed region contrast (see Appendix~\ref{appendix:jupiter_transit}).

Empirical models predict lower spot temperatures than our estimates for eclipsed active regions. One such model was derived by \citet{berdyugina2005}. To estimate spot temperatures in their sample, they used different methods, including Doppler imaging, light curve modelling, molecular bands, and atomic line-depth ratios. \citet{herbst2021} updated this empirical relation upon extending the original sample of \citet{berdyugina2005}. As an example, for HD~189733, Eq.~6 of \citet{herbst2021} predicts a temperature difference of $\Delta T = 578 \pm 467$~K between the umbra and the unspotted photosphere, which is a factor of five bigger than the maximum temperature difference calculated from the contrast of an active region ($\Delta T \approx 120$~K). However, both \citet{berdyugina2005} and \citet{herbst2021} advise using their relations with caution, since e.g., they cannot reproduce the solar spot temperature contrasts.

All planets with candidate spot occultations had orbital periods shorter than 9 days. These planets have relatively large radii compared to their stars, ranging from $0.045$ to $0.18$. The signal-to-noise ratio distribution of the transit light curves for planets with spot eclipse detections and those without were reasonably similar.

\begin{figure}[]
    \centering
    \includegraphics[width=.5\textwidth]{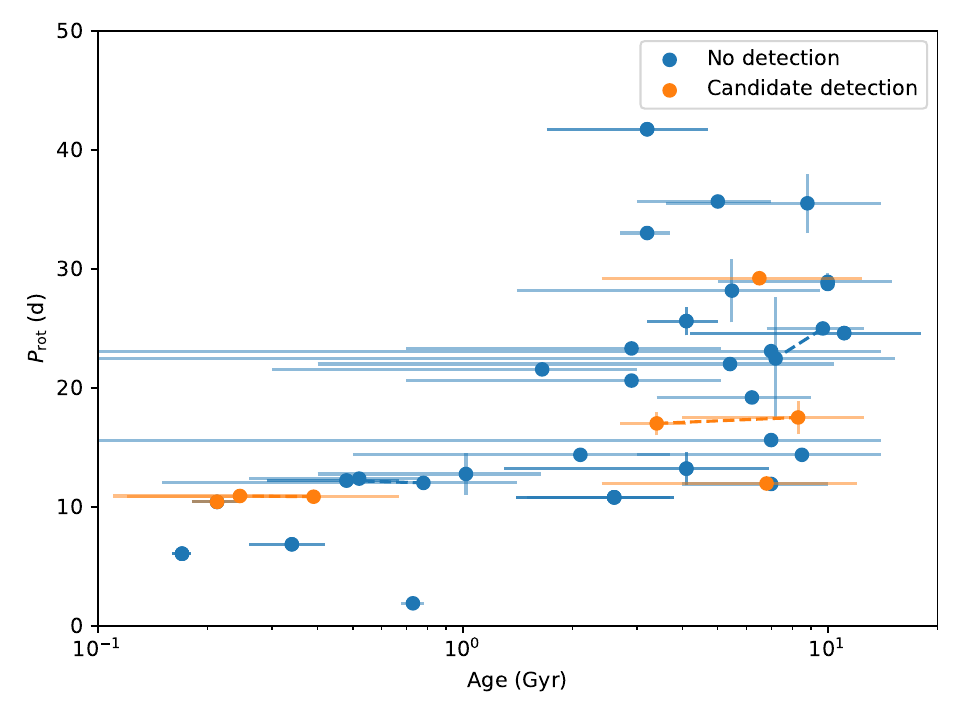}
      \caption{Comparison of ages and rotation periods of the stars with spot eclipse candidates (orange) and stars where no spot eclipse was detected in our analysis (blue).}
         \label{fig:prot_age}
   \end{figure}

Of the $89$ stars in our sample, $80$ had age estimates and $41$ had rotation periods listed in the NASA Exoplanet Archive, with $38$ having both. Since the archive may lack some published values, we reviewed the literature for stars with spot eclipse candidates. Of the six such planets, five had age estimates and all six had rotation periods. Two stars with spot eclipse candidates are younger than 1~Gyr (Kepler-411 and TOI-1268, $212 \pm 31$~Myr and $245 \pm 135$~Myr, respectively), both have $\sim 10$~d rotational periods. Fig.~\ref{fig:prot_age} compares the ages and rotation periods of the stars with spot eclipse candidates and those where no eclipse signals were found in the analysis. The figure only shows those stars whose ages and rotation periods were derived simultaneously. Different measurements for the same star are connected by dashed lines.

It is worth comparing the scale of the transited features with the resolution of other methods to resolve the surfaces of late-type stars. One common method for recovering the brightness distribution of late-type stars is Doppler imaging, which relies on a time-series of spectroscopic measurements of a star \citep[see e.g.,][]{piskunov2002}. Coupled with polarimetric measurements, the large scale magnetic field topology of the star may also be recovered, which can serve as a basis for the reconstruction of the surface brightness distribution. Depending on the phase coverage and the projected rotational velocity ($v \sin i$), these methods can reveal surface details as small as $5$~degrees \citep{carroll2012}.

While Doppler imaging is limited in resolution due to requirements on stellar rotational velocity, and the resolution and signal-to-noise ratio of the spectra, transit mapping has a higher potential to discern the smallest dark features of the stellar surface. We found that the main limiting factor for the detection of small spots is the signal-to-noise ratio of the data. The smallest spot in our sample had a radius of $1.6$~degrees (HAT-P-11).
Inverse photometric modelling of rotational signals may discern dark features on similar size scales, but fails to recover spot contrasts. A small, high contrast spot would produce a similar signal as a larger, low contrast region \citep[see e.g.,][]{tuomi2024}.

Our sample contained 8 K-dwarfs and 4 M-dwarfs observed by the Kepler or K2 missions, and 54 K-dwarfs and 24 M-dwarfs observed by TESS. Kepler and TESS both provided light curves for the candidate spot occultation detections of three stars (Kepler: HAT-P-11, Kepler-411, and WASP-107, TESS: HD~189733, TOI-1268, and TOI-3884). Of the stars with candidate spot eclipses, only TOI-3884 was an M-dwarf, while the rest were K-dwarfs.

The spot detection frequency of Kepler was $37.5\%$ for K-dwarfs, while for TESS, it was $3.7\%$ and $4.2\%$, for K- and M-dwarfs, respectively. Using a set of simulated TESS light curves, \citet{tregloan2021} predicted that spot detection frequencies for K- and M-dwarfs in the TESS sample would be $10.66\% \pm 0.41\%$ and $17.26\% \pm 0.48\%$, respectively. The difference between their results and those presented in this analysis may be due to the detection criteria they applied. The authors compared flux deficits to the root mean square scatter without considering event length or phase. With TESS expected to find $1700$~planets \citep{sullivan2015}, a larger sample is needed for more robust spot detection statistics.

Photometric variations during transits could, in theory, originate from exomoons transiting the disk of their planets. To test this, we calculated the minimum duration of an exomoon transit in front of the disk of its planet. Following \citet{dobos2021}, we approximate the semi-major axis of the innermost stable orbit of a moon as $2 R_p$ with circular orbits.  For planets with spot eclipse candidates, predicted exomoon transit durations range from $\sim 3230$~s (HAT-P-11~b) to $\sim 15\,000$~s (WASP-107~b). Only one spot eclipse candidate (HAT-P-11~b, transit $-369$) exceeded this threshold. Additionally, planets with multiple spot eclipses show varying signal amplitudes, which is impossible from exomoon transits.

Another systematic search for spot occultations was conducted by \citet{baluev2021}. They analysed spot- and facula-induced modulations in $1598$ transit light curves from $26$ targets, collected through a network of amateur and professional observatories. Their spot model was a simple Gaussian profile, corresponding to the extreme case of $t_{\rm in} = t_{\rm out}$ in our Eq.~\ref{eq:spot_occultation}. They applied an iterative algorithm to detect Gaussian anomalies, retaining only statistically significant and well-justified features. Overall, they reported $38$ potential occultations of dark regions.

Our spot contrast results align reasonably with most of \citet{baluev2021}. However, due to their definition of occultation length (the $\sigma$ of the Gaussian anomaly), their occultations occur on a shorter timescale than those identified in this analysis. Multiplying their occultation lengths by a factor of 2 reduces these differences, but several of our candidate eclipses still occur on longer timescales compared to those of \citet{baluev2021}, even after the adjustment.

\citet{baluev2021} reported spot eclipse detections for $20$ of $26$ stars, while we found occultation candidates for $6$ of $89$ stars. This results in a detection rate $11$~times higher in their study ($76.9\%$) compared to ours ($6.7\%$). Among the $5$ stars analysed in both studies (HAT-P-12, HD~189733, Qatar-4, TrES-1, and WASP-2), we identified candidate occultation events only for HD~189733. Several factors may explain the differences in detection statistics between the two studies. Firstly, none of the observations of \citet{baluev2021} were conducted using the space telescopes employed in our study. Secondly, improper normalisation of some out-of-transit light curve segments in their dataset could have introduced artefacts into the transit signals. These artefacts might cause deviations from the quadratic limb-darkening law assumed in their analysis, resulting in the appearance of spot-like features. Furthermore, the dataset of \citet{baluev2021} includes spot occultations occurring during the ingress or egress phases of transits. While such events are also allowed in our analysis, due to their lower expected amplitudes, they are unlikely to be confirmed by our detection criteria.

\begin{figure}[]
    \centering
    \includegraphics[width=.5\textwidth]{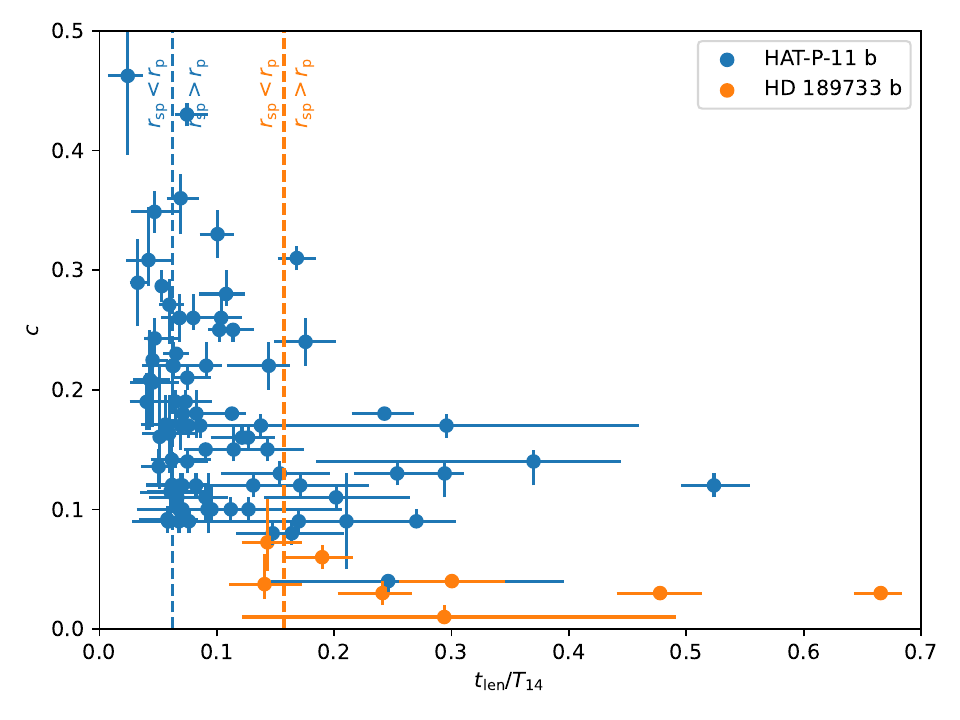}
      \caption{The relative durations $t_{\rm len}/T_{14}$ and contrasts $c$ of dark region occultation candidates are shown for the two planets with the most events. Dashed vertical lines represent durations where the angular radius of the dark regions $r_{\rm sp}$ matches that of the planet $r_{\rm p}$. For durations below this threshold, the contrasts have been adjusted using Eq.~\ref{eq:contrast_correction}.}
         \label{fig:tlen_vs_c}
   \end{figure}

Only two objects, HAT-P-11~b and HD~189733~b, had more than three spot occultation candidates in our sample. Fig.~\ref{fig:tlen_vs_c} compares the durations of spot occultation events and contrasts of the eclipsed regions for these two planets. The eclipsed dark regions on HAT-P-11 are generally smaller and darker compared to those on HD~189733. This difference in spot properties may be attributed to different planet-star radius ratios and orbital alignments. The radius ratios for HAT-P-11~b and HD~189733~b are $r = 0.06$ and $r = 0.15$, respectively. HAT-P-11~b, being smaller, can completely overlap with individual dark regions on the stellar surface, while the larger HD~189733~b may cover both bright and dark regions, reducing the contrast of the observed signal. Additionally, HAT-P-11~b is orbitally misaligned, allowing it to cover a wide range of stellar latitudes. No such misalignment is observed for HD~189733~b. If active regions are spread longitudinally on these stars, HAT-P-11~b would briefly transit these regions, while HD~189733~b would cover the same latitudes throughout its transit.

We found that by far the most false positives came from the varying transit depth from stars with low signal-to-noise observations. Outlying flux observations may also cause high enough Bayes factors for spot detections. Both these cases can be easily refuted with visual inspection.

The frequency of spot eclipses in a planet’s transits allows us to estimate the mean spot number $N_{\rm S}$. Combined with spot sizes, we may estimate the range of possible spot filling factors $f_{\rm S}$ of each star. Table~\ref{tab:spot_detections} lists the mean spot numbers, and the range of spot filling factors $[f_{\rm S,min}, f_{\rm S,max}]$ for each star where a spot eclipse candidate was found in our sample. Spot filling factors are derived from the fractional angular areas of the smallest and largest eclipsed spots, multiplied by the mean number of spots.
The spot numbers and spot filling factors are approximated with the assumption that the spots are distributed uniformly on the stellar surface. 
At least for TOI-3884, we have good reason to believe that this does not hold, as its planet is probably repeatedly eclipsing the polar spot of the star. Our analysis also failed to pick up numerous spot eclipse events of the same spot, due to the noisiness of the data. Because of this, we advise caution when handling the spot occurrence statistics of this star.

The upper limits for spot filling factors we observed for the stars range between $0.9\%$ and $17.3\%$.
For every star in our sample, the spot filling factors have remained at least a factor of two lower than what is inferred for young K- and M-dwarfs based on studies of TiO absorption bands \citep{oneal2004}.
This discrepancy may arise because transit mapping is more sensitive to the stellar centre or because smaller, lower-contrast spots remain undetected because of observational noise.

\begin{table}
\caption{Number of analysed transits $n_{\rm tr}$, mean spot numbers $N_{\rm S}$, and the range of possible spot covering fractions $[f_{\rm S,min}, f_{\rm S,max}]$ based on the angular sizes of spot candidates for each star in the analysed sample.} 
\label{tab:spot_detections}      
\centering                          
\begin{tabular}{lrrrr}        
\hline\hline                 
\multicolumn{1}{c}{Star} & \multicolumn{1}{c}{$n_{\rm tr}$} & \multicolumn{1}{c}{$N_{\rm S}$} & \multicolumn{1}{c}{$f_{\rm S,min}$} & \multicolumn{1}{c}{$f_{\rm S,max}$}  \\     
\hline                        
HAT-P-11\tablefootmark{a} & 183 & $12.92$ & 0.0025 & 0.1732 \\
HD 189733 & 21 & $3.79$ & 0.0058 & 0.1333 \\
Kepler-411\tablefootmark{b} & 45 & $2.46$ & 0.0009 & 0.0089 \\
TOI-1268 & 14 & $0.63$ & 0.0281 & 0.0281 \\
TOI-3884 & 8 & $2.38$ & 0.0603 & 0.1545 \\
WASP-107 & 9 & $3.73$ & 0.0092 & 0.1081 \\
\hline
\end{tabular}
\tablefoot{
\tablefoottext{a}{Spot statistics only based on Kepler observations.}
\tablefoottext{b}{Spot statistics only based on transits of planet~c.}
}
\end{table}

\begin{table}
\caption{Expected maximum flux change from co-rotating starspots $\Delta F_{\rm spot}$ compared to the amplitude of the rotational modulation observed around the transit $A_{\rm phot}$. The data column tells whether amplitudes were derived from SAP or PDC light curves.}             
\label{tab:sap_comparison}      
\centering                          
\begin{tabular}{lrrrr}        
\hline\hline                 
\multicolumn{1}{c}{Planet} & \multicolumn{1}{c}{Transit} & \multicolumn{1}{c}{Data} & \multicolumn{1}{c}{$A_{\rm phot}/10^{-5}$ } & \multicolumn{1}{c}{$\Delta F_{\rm spot}/10^{-5}$}  \\     
\hline                        
HD189733 b & 2161 & SAP &1454	&   $  16 \pm    9$ \\
HD189733 b & 2165 & SAP &1454	&   $  34 \pm   18$ \\
HD189733 b & 2167 & SAP &1454	&   $  64 \pm   20$ \\
HD189733 b & 2168 & SAP &1260	&   $ 113 \pm   18$ \\
HD189733 b & 2170 & SAP &882	&   $  33 \pm   11$ \\
HD189733 b & 2317 & SAP &954	&   $  26 \pm   11$ \\
HD189733 b & 2323 & SAP &956	&   $ 212 \pm   14$ \\
HD189733 b & 2326 & SAP &832	&   $  28 \pm   24$ \\
Kepler-411 c & $116$ & SAP &1054&   $  35 \pm   24$ \\
Kepler-411 c & $117$ & SAP &1288&   $ 149 \pm   64$ \\
Kepler-411 c & $121$ & SAP &175 &   $  27 \pm   17$ \\
TOI-1268 b & 3 & SAP &1663	&   $1432 \pm  370$ \\
TOI-3884 b & $-16$ & PDC &1468	&   $4047 \pm 2263$ \\
TOI-3884 b & $-15$ & PDC &1189	&   $1984 \pm 1078$ \\
TOI-3884 b & $1$ & PDC &1103	&   $2782 \pm 1262$ \\
WASP-107 b & $4$ & PDC &555	&   $ 122 \pm   21$ \\
WASP-107 b & $6$ & PDC &793	&   $  79 \pm   15$ \\
WASP-107 b & $11$ & PDC &408	&   $ 127 \pm   68$ \\

\hline
\end{tabular}
\end{table}

We calculated the maximum brightness variation each spot could induce per stellar rotation and compared it with flux variations around transit events where spots were detected. We extracted the observations in a 30~d window centred around the spot eclipse event, and smoothened the observations by taking the running mean of every 20th flux measurement for 120~s observations, and every 40th flux measurement for 60~s observations. Flares were omitted from the data after visual vetting. We used SAP fluxes for this comparison, unless they were contaminated by instrumental drift. We preferred SAP fluxes to PDSCAP fluxes due to the fact that stellar variability over timescales longer than around $10$~days may get significantly dampened by the data conditioning pipeline \citep[see e.g.,][]{gilliland2015}. Table~\ref{tab:sap_comparison} compares the amplitude of brightness variations $A_{\rm phot}$ in the Kepler and TESS observations around each spot occultation candidate and the maximum possible flux change $\Delta F_{\rm spot}$ the dark region can cause on the star. This comparison is presented separately for the spot candidates of HAT-P-11~b, in Table~\ref{tab:sap_comparison11} of the Appendix. The estimated flux variations from all non-axisymmetric spots in our sample were consistent with the observed flux changes over the stellar rotation.

\section{Conclusions}

We performed a uniform analysis of possible spot occultation events by exoplanets in the Kepler and TESS high cadence observations of K- and M-dwarfs. We analysed 3273 transits from 99 planets in total, and found strong evidence for spot eclipses on 105 occasions, by $6$~planets.

We report new spot eclipse candidates for HD~189733~b in recent TESS data, confirming past HST detections \citep{pont2007, sing2011}, and an individual candidate for TOI-1268~b. Our analysis supports previous detections for HAT-P-11~b \citep{morris2017} and WASP-107~b \citep{dai2017} and recovers some spot eclipses from Kepler-411~c \citep{araujo2021} and TOI-3884~b \citep{almenara2022, libby2023}. Differences in detection criteria and noise treatment likely explain some non-detections.

The candidate eclipse durations range from approximately $200$ to $5300$ seconds, with minimum radii of the dark regions spanning $1.6^\circ$ to $29.5^\circ$. Kepler's spot detection frequency for K-dwarfs was $37.5\%$, while for TESS, it was $3.7\%$ and $4.2\%$ for K- and M-dwarfs, respectively. The observed upper limits for spot filling factors range from $0.9\%$ to $17.3\%$.

In this work, we have introduced a new modelling framework to analyse large quantities of planetary transit observations, aimed at building robust spot statistics. As the number of known exoplanets around late-type stars is expected to grow, the likelihood of detecting spot occultation events will also increase. In the process, we will eventually get reliable statistics of the smallest detectable spots on other stars, and, potentially improve the assumptions and understanding of the magnetic activity of late-type stars.

\begin{acknowledgements}
We are thankful to the anonymous referee and to Katalin Ol\'ah for valuable feedback. AH and MT gratefully acknowledge support from the Jenny and Antti Wihuri Foundation. The authors acknowledge the Research Council of Finland project SOLSTICE (decision No.~324161). The authors wish to acknowledge CSC – IT Center for Science, Finland, for computational resources and support. This research has made use of NASA's Astrophysics Data System. This research has made use of adstex (\url{https://github.com/yymao/adstex}).
\end{acknowledgements}

%
%

\bibliography{aanda.bib}

\begin{appendix}
\section{Target list}

In Table~\ref{tab:analysed_sample}, we present the number of analysed transits from each mission, and the orbital and planetary parameters that were used to extract the transit light curves and for initial values in the posterior sampling. Planets marked with an asterisk are listed with orbital periods $P$ and times of conjunction $T_0$ that have been recalculated for this work.

\onecolumn{
\begin{longtable}{lrrrrrrrrrc}
\caption{\label{tab:analysed_sample} Sample properties for the analysed planets: number of analysed transit light curves ($n_{\rm tr}$) from the Kepler (K), K2 (K2), and TESS (T) missions, orbital period ($P$), time of conjuction ($T_0$), transit length ($T_{14}$), planet--star radius ratio ($r$), inclination of planetary orbit ($i$), stellar radius ($R_*$) and the references (Ref.) for the listed values. Planets marked with an asterisk have updated $P$ and $T_0$ parameters, estimated in this work.}\\
\hline\hline
\multicolumn{1}{c}{\begin{tabular}[c]{@{}c@{}}Planet\end{tabular}} & \multicolumn{1}{c}{\begin{tabular}[c]{@{}c@{}}$n_{\mathrm{tr}}$\\ (K)\end{tabular}} & \multicolumn{1}{c}{\begin{tabular}[c]{@{}c@{}}$n_{\mathrm{tr}}$\\ (K2)\end{tabular}} & \multicolumn{1}{c}{\begin{tabular}[c]{@{}c@{}}$n_{\mathrm{tr}}$\\ (T)\end{tabular}} & \multicolumn{1}{c}{\begin{tabular}[c]{@{}c@{}}$P$ \\ (d)\end{tabular}} & \multicolumn{1}{c}{\begin{tabular}[c]{@{}c@{}}$T_0$\\ (BJD-2450000)\end{tabular}} & \multicolumn{1}{c}{\begin{tabular}[c]{@{}c@{}}$T_{14}$\\ (h)\end{tabular}} & \multicolumn{1}{c}{$r$} & \multicolumn{1}{c}{\begin{tabular}[c]{@{}c@{}}$i$\\ ($^\circ$)\end{tabular}}  & \multicolumn{1}{c}{\begin{tabular}[c]{@{}c@{}}$R_*$\\ ($R_{\odot}$)\end{tabular}}  & \multicolumn{1}{c}{\begin{tabular}[c]{@{}c@{}}Ref.\\ \end{tabular}} \\
\hline
\endfirsthead
\caption{continued.}\\
\hline\hline
\multicolumn{1}{c}{\begin{tabular}[c]{@{}c@{}}Planet\end{tabular}} & \multicolumn{1}{c}{\begin{tabular}[c]{@{}c@{}}$n_{\mathrm{tr}}$\\ (K)\end{tabular}} & \multicolumn{1}{c}{\begin{tabular}[c]{@{}c@{}}$n_{\mathrm{tr}}$\\ (K2)\end{tabular}} & \multicolumn{1}{c}{\begin{tabular}[c]{@{}c@{}}$n_{\mathrm{tr}}$\\ (T)\end{tabular}} & \multicolumn{1}{c}{\begin{tabular}[c]{@{}c@{}}$P$ \\ (d)\end{tabular}} & \multicolumn{1}{c}{\begin{tabular}[c]{@{}c@{}}$T_0$\\ (BJD-2450000)\end{tabular}} & \multicolumn{1}{c}{\begin{tabular}[c]{@{}c@{}}$T_{14}$\\ (h)\end{tabular}} & \multicolumn{1}{c}{$r$} & \multicolumn{1}{c}{\begin{tabular}[c]{@{}c@{}}$i$\\ ($^\circ$)\end{tabular}}  & \multicolumn{1}{c}{\begin{tabular}[c]{@{}c@{}}$R_*$\\ ($R_{\odot}$)\end{tabular}}  & \multicolumn{1}{c}{\begin{tabular}[c]{@{}c@{}}Ref.\\ \end{tabular}} \\
\hline
\endhead
\hline
\endfoot
AU Mic b & 0 & 0 & 1 & 8.4631 & 8330.3905 & 3.56 &0.051& 89.18 & 0.82 & 1, 2, 3 \\
GJ 143 b & 0 & 0 & 6 & 35.6125 & 8385.9250 & 3.23 &0.035& 89.33 & 0.69 & 4 \\
GJ 3470 b & 0 & 0 & 18 & 3.3366 & 5953.6630 & 1.92 &0.076& 89.13 & 0.48 & 5, 6, 7 \\
GJ 9827 b & 0 & 2 & 0 & 1.2090 & 7738.8259 & 1.26 &0.024& 86.07 & 0.60 & 8, 9, 10 \\
HAT-P-11 b & 183 & 0 & 31 & 4.8878 & 7631.4409 & 2.36 &0.058& 89.82 & 0.75 & 11, 12, 13, 14 \\
HAT-P-12 b & 0 & 0 & 14 & 3.2131 & 8943.1819 & 2.17 &0.139& 88.24 & 0.70 & 15, 16, 17, 18, 19 \\
HAT-P-17 b & 0 & 0 & 6 & 10.3385 & 8719.4753 & 3.91 &0.124& 89.20 & 0.84 & 11, 16, 20, 21 \\
HAT-P-18 b & 0 & 0 & 7 & 5.5080 & 9033.3174 & 2.64 &0.132& 88.66 & 0.75 & 15, 16, 21, 22, 23 \\
HAT-P-19 b & 0 & 0 & 6 & 4.0088 & 6935.5755 & 2.74 &0.138& 88.20 & 0.82 & 11, 16, 21, 23, 24 \\
HAT-P-20 b & 0 & 0 & 35 & 2.8753 & 6233.9298 & 1.88 &0.155& 86.80 & 0.69 & 11, 16, 21, 25, 26 \\
HAT-P-26 b & 0 & 0 & 3 & 4.2345 & 6545.3612 & 2.52 &0.074& 88.60 & 0.79 & 11, 16, 27, 28 \\
HAT-P-3 b & 0 & 0 & 19 & 2.8997 & 4978.4910 & 1.85 &0.106& 86.31 & 0.83 & 11, 16, 17, 29, 30 \\
HAT-P-68 b & 0 & 0 & 35 & 2.2984 & 6614.2036 & 2.09 &0.164& 88.73 & 0.67 & 31 \\
HATS-2 b & 0 & 0 & 48 & 1.3541 & 8582.9600 & 1.89 &0.134& 87.20 & 0.90 & 15, 16, 21, 32 \\
HATS-22 b & 0 & 0 & 13 & 4.7228 & 8580.4362 & 2.19 &0.140& 87.96 & 0.69 & 15, 16, 33 \\
HATS-6 b & 0 & 0 & 22 & 3.3253 & 9227.4723 & 2.05 &0.167& 88.21 & 0.57 & 15, 16, 34 \\
HATS-71 b & 0 & 0 & 26 & 3.7955 & 9111.3229 & 2.01 &0.220& 88.82 & 0.48 & 15, 35 \\
HD 189733 b & 0 & 0 & 21 & 2.2186 & 4632.1905 & 1.84 &0.150& 85.69 & 0.76 & 11, 30, 36, 37 \\
HD 207496 b & 0 & 0 & 5 & 6.4410 & 8658.7898 & 1.76 &0.027& 88.79 & 0.77 & 38 \\
HD 219134 b* & 0 & 0 & 7 & 3.0929 & 8765.9545 & 0.93 &0.018& 85.01 & 0.77 & 37, 39 \\
HD 219134 c* & 0 & 0 & 8 & 6.7650 & 8766.1693 & 1.62 &0.018& 87.30 & 0.77 & 37, 39 \\
HD 73583 b & 0 & 0 & 4 & 6.3980 & 9240.6702 & 2.07 &0.035& 89.43 & 0.65 & 40 \\
HD 95338 b & 0 & 0 & 1 & 55.0870 & 8585.2795 & 5.71 &0.041& 89.57 & 0.87 & 16, 41, 42 \\
HD 97658 b & 0 & 0 & 2 & 9.4893 & 7785.2018 & 2.76 &0.027& 89.05 & 0.73 & 11, 16, 37, 43 \\
HIP 65 A b* & 0 & 0 & 83 & 0.9810 & 8354.5514 & 0.79 &0.287& 77.18 & 0.73 & 44, 45 \\
K2-110 b & 0 & 1 & 0 & 13.8637 & 7275.3299 & 3.22 &0.034& 89.35 & 0.71 & 8, 46 \\
K2-146 c & 0 & 12 & 0 & 3.9663 & 8299.6019 & 1.51 &0.068& 87.30 & 0.35 & 46, 47 \\
K2-25 b & 0 & 0 & 5 & 3.4846 & 7620.1105 & 0.76 &0.107& 87.16 & 0.29 & 48, 49, 50 \\
K2-29 b & 0 & 0 & 13 & 3.2588 & 8508.0943 & 2.20 &0.129& 86.97 & 0.86 & 46, 48, 51, 52, 53 \\
K2-295 b & 0 & 0 & 12 & 4.0249 & 7395.4139 & 2.51 &0.132& 89.30 & 0.67 & 46, 54 \\
KPS-1 b & 0 & 0 & 53 & 1.7063 & 8883.6684 & 1.68 &0.104& 83.20 & 0.91 & 15, 55 \\
Kepler-210 b & 2 & 0 & 0 & 2.4532 & 5003.8954 & 1.67 &0.039& 87.23 & 0.64 & 12, 56, 57 \\
Kepler-210 c & 101 & 0 & 0 & 7.9725 & 5004.5833 & 2.98 &0.055& 87.26 & 0.64 & 12, 56, 57 \\
Kepler-411 b & 1 & 0 & 0 & 3.0052 & 5135.3874 & 2.07 &0.027& 87.40 & 0.82 & 12, 58 \\
Kepler-411 c & 45 & 0 & 0 & 7.8344 & 4968.2224 & 2.89 &0.045& 88.61 & 0.82 & 12, 58 \\
Kepler-411 d & 2 & 0 & 0 & 58.0204 & 4984.8484 & 5.35 &0.037& 89.43 & 0.82 & 12, 58 \\
Kepler-425 b & 77 & 0 & 0 & 3.7970 & 4966.5088 & 2.26 &0.109& 88.65 & 0.83 & 12, 56, 57 \\
Kepler-45 b & 286 & 0 & 0 & 2.4552 & 5003.8218 & 1.73 &0.171& 89.35 & 0.61 & 12, 56, 57 \\
Kepler-75 b & 43 & 0 & 0 & 8.8849 & 5740.4403 & 2.07 &0.114& 89.89 & 0.98 & 12, 56, 57 \\
Kepler-94 b & 339 & 0 & 0 & 2.5081 & 4968.0004 & 1.11 &0.039& 86.61 & 0.74 & 12, 56, 57 \\
L 98-59 c & 0 & 0 & 54 & 3.6907 & 8367.2738 & 1.35 &0.039& 88.11 & 0.31 & 59, 60, 61 \\
L 98-59 d & 0 & 0 & 64 & 7.4507 & 8362.7398 & 0.84 &0.046& 88.45 & 0.31 & 59, 60, 61 \\
LHS 1140 b & 0 & 0 & 2 & 24.7369 & 8399.9303 & 2.06 &0.074& 89.88 & 0.21 & 62, 63 \\
LP 714-47 b & 0 & 0 & 12 & 4.0520 & 9196.1149 & 1.56 &0.075& 87.30 & 0.58 & 15, 16, 64 \\
NGTS-1 b* & 0 & 0 & 15 & 2.6462 & 8469.8515 & 1.23 &0.190& 85.27 & 0.58 & 15, 65 \\
NGTS-5 b & 0 & 0 & 2 & 3.3570 & 7740.3526 & 2.21 &0.158& 86.60 & 0.74 & 66 \\
Qatar-4 b & 0 & 0 & 12 & 1.8054 & 7762.3441 & 2.08 &0.138& 87.50 & 0.85 & 11, 16, 67 \\
Qatar-6 b & 0 & 0 & 5 & 3.5062 & 7836.6256 & 1.59 &0.151& 86.01 & 0.72 & 11, 68 \\
Qatar-9 b & 0 & 0 & 30 & 1.5408 & 8871.8006 & 1.87 &0.149& 89.23 & 0.70 & 11, 69 \\
TOI-1130 b* & 0 & 0 & 8 & 4.0653 & 9037.9399 & 2.30 &0.049& 87.98 & 0.70 & 70 \\
TOI-1130 c* & 0 & 0 & 4 & 8.3495 & 9042.0005 & 2.02 &0.218& 87.43 & 0.70 & 70 \\
TOI-1268 b & 0 & 0 & 14 & 8.1577 & 8703.5895 & 4.00 &0.089& 89.55 & 0.86 & 71, 72 \\
TOI-1278 b* & 0 & 0 & 6 & 14.4751 & 8711.9593 & 1.62 &0.197& 88.30 & 0.86 & 71, 73 \\
TOI-1728 b* & 0 & 0 & 14 & 3.4914 & 8843.2757 & 1.96 &0.071& 88.31 & 0.86 & 15, 74 \\
TOI-1759 b & 0 & 0 & 3 & 18.8502 & 8745.4654 & 3.23 &0.048& 89.72 & 0.60 & 75 \\
TOI-181 b & 0 & 0 & 14 & 4.5320 & 8358.1200 & 1.99 &0.085& 88.28 & 0.74 & 76 \\
TOI-1899 b* & 0 & 0 & 4 & 29.0217 & 8715.7836 & 4.67 &0.194& 89.77 & 0.74 & 77 \\
TOI-2076 c & 0 & 0 & 1 & 21.0154 & 9274.0840 & 4.19 &0.042& 89.84 & 0.77 & 78, 79 \\
TOI-2076 d & 0 & 0 & 1 & 35.1254 & 8938.2915 & 3.05 &0.038& 88.61 & 0.77 & 78, 79 \\
TOI-2202 b* & 0 & 0 & 6 & 11.9140 & 9089.5528 & 4.01 &0.126& 88.40 & 0.77 & 80 \\
TOI-270 c & 0 & 0 & 20 & 5.6606 & 8389.5028 & 1.66 &0.057& 89.36 & 0.38 & 81, 82 \\
TOI-270 d & 0 & 0 & 4 & 11.3801 & 9129.3487 & 2.15 &0.053& 89.68 & 0.38 & 81, 82, 83 \\
TOI-3235 b & 0 & 0 & 9 & 2.5926 & 9690.0017 & 1.48 &0.283& 88.14 & 0.37 & 84 \\
TOI-3629 b & 0 & 0 & 5 & 3.9366 & 9662.1080 & 2.19 &0.125& 88.74 & 0.61 & 85 \\
TOI-3714 b & 0 & 0 & 10 & 2.1548 & 9687.3652 & 1.63 &0.206& 87.83 & 0.50 & 85 \\
TOI-3757 b & 0 & 0 & 11 & 3.4388 & 8838.7715 & 1.92 &0.177& 86.76 & 0.62 & 86 \\
TOI-3785 b & 0 & 0 & 1 & 4.6747 & 8861.4955 & 1.70 &0.096& 88.10 & 0.50 & 87 \\
TOI-3884 b & 0 & 0 & 8 & 4.5446 & 9642.8631 & 1.65 &0.181& 90.10 & 0.30 & 88 \\
TOI-4600 b & 0 & 0 & 3 & 82.6869 & 9750.1421 & 7.54 &0.077& 89.76 & 0.81 & 89 \\
TOI-4600 c & 0 & 0 & 1 & 482.8191 & 9751.6008 & 11.14 &0.107& 89.90 & 0.81 & 89 \\
TOI-4860 b & 0 & 0 & 14 & 1.5228 & 9832.6414 & 1.24 &0.220& 88.87 & 0.35 & 90 \\
TOI-519 b & 0 & 0 & 54 & 1.2652 & 9013.1530 & 1.24 &0.304& 88.90 & 0.35 & 85, 91 \\
TOI-5293 A b & 0 & 0 & 6 & 2.9303 & 9448.9148 & 1.94 &0.210& 88.80 & 0.52 & 92 \\
TOI-674 b* & 0 & 0 & 45 & 1.9772 & 8544.5251 & 1.04 &0.114& 87.21 & 0.52 & 93 \\
TOI-700 c & 0 & 0 & 8 & 16.0511 & 8821.6218 & 1.42 &0.057& 88.90 & 0.42 & 94 \\
TOI-836 c & 0 & 0 & 5 & 8.5954 & 8599.7623 & 2.49 &0.036& 88.70 & 0.67 & 95 \\
TrES-1 b & 0 & 0 & 27 & 3.0301 & 6168.3959 & 2.55 &0.138& 90.00 & 0.81 & 11, 16, 21, 30, 96 \\
WASP-10 b & 0 & 0 & 8 & 3.0927 & 6188.7535 & 2.22 &0.161& 88.81 & 0.78 & 11, 16, 21, 97, 98 \\
WASP-105 b & 0 & 0 & 8 & 7.8729 & 9111.5293 & 3.72 &0.113& 89.70 & 0.90 & 15, 99 \\
WASP-107 b & 0 & 9 & 0 & 5.7215 & 7584.3295 & 2.74 &0.144& 89.56 & 0.66 & 11, 99, 100, 101 \\
WASP-11 b & 0 & 0 & 10 & 3.7225 & 6263.5689 & 2.60 &0.142& 89.03 & 0.74 & 11, 16, 21, 101, 102 \\
WASP-132 b & 0 & 0 & 9 & 7.1335 & 8609.9896 & 3.18 &0.125& 89.51 & 0.74 & 15, 103, 104 \\
WASP-139 b & 0 & 0 & 8 & 5.9243 & 8387.5716 & 2.89 &0.101& 88.90 & 0.85 & 11, 16, 103, 104 \\
WASP-140 b & 0 & 0 & 19 & 2.2360 & 8412.6973 & 1.53 &0.142& 83.30 & 0.87 & 11, 16, 103, 104 \\
WASP-144 b & 0 & 0 & 19 & 2.2783 & 9071.0588 & 1.95 &0.113& 86.90 & 0.81 & 15, 105 \\
WASP-145 A b & 0 & 0 & 22 & 1.7690 & 9037.7733 & 0.98 &0.136& 83.30 & 0.68 & 11, 15, 105 \\
WASP-156 b & 0 & 0 & 29 & 3.8362 & 9169.8600 & 2.41 &0.067& 89.10 & 0.76 & 15, 106 \\
WASP-160 B b & 0 & 0 & 12 & 3.7685 & 9203.8382 & 2.84 &0.129& 88.97 & 0.87 & 11, 107 \\
WASP-162 b & 0 & 0 & 3 & 9.6247 & 8548.3525 & 4.26 &0.093& 89.30 & 1.11 & 11, 105, 107 \\
WASP-177 b & 0 & 0 & 11 & 3.0717 & 9450.3674 & 1.61 &0.182& 84.14 & 0.89 & 11, 107, 108 \\
WASP-2 b & 0 & 0 & 4 & 2.1522 & 5381.8501 & 1.79 &0.128& 84.49 & 0.84 & 11, 30, 36 \\
WASP-43 b & 0 & 0 & 78 & 0.8135 & 8555.8057 & 1.23 &0.162& 82.33 & 0.60 & 15, 16, 21, 109 \\
WASP-45 b & 0 & 0 & 20 & 3.1261 & 6497.8858 & 1.68 &0.114& 85.02 & 0.94 & 11, 36, 110 \\
WASP-59 b & 0 & 0 & 3 & 7.9193 & 9830.3490 & 2.51 &0.139& 89.27 & 0.76 & 16, 18, 21 \\
WASP-69 b & 0 & 0 & 3 & 3.8681 & 7176.1779 & 2.16 &0.139& 86.71 & 0.81 & 16, 18, 21, 48, 111 \\
WASP-80 b & 0 & 0 & 5 & 3.0679 & 6824.8884 & 2.11 &0.171& 89.92 & 0.57 & 11, 16, 21, 112, 113 \\
WASP-84 b & 0 & 0 & 4 & 8.5235 & 7956.7119 & 2.76 &0.128& 88.29 & 0.77 & 114 \\
WASP-91 b & 0 & 0 & 33 & 2.7986 & 9085.1043 & 2.34 &0.121& 86.80 & 0.86 & 15, 99 \\
Wolf 503 b & 0 & 7 & 0 & 6.0013 & 8191.3615 & 1.33 & 0.028 & 89.87 & 0.69 & 8, 115
\end{longtable}
\tablebib{
(1)~\citet{https://ui.adsabs.harvard.edu/abs/2023MNRAS.525..455D/abstract};
(2)~\citet{https://ui.adsabs.harvard.edu/abs/2022MNRAS.512.3060Z/abstract};
(3)~\citet{https://ui.adsabs.harvard.edu/abs/2022AJ....163..147G/abstract};
(4)~\citet{https://ui.adsabs.harvard.edu/abs/2019ApJ...875L...7D/abstract};
(5)~\citet{https://ui.adsabs.harvard.edu/abs/2019AJ....157...97K/abstract};
(6)~\citet{https://ui.adsabs.harvard.edu/abs/2014MNRAS.443.1810B/abstract};
(7)~\citet{https://ui.adsabs.harvard.edu/abs/2016MNRAS.463.2574A/abstract};
(8)~\citet{https://ui.adsabs.harvard.edu/abs/2023arXiv230405773B/abstract};
(9)~\citet{https://ui.adsabs.harvard.edu/abs/2019MNRAS.484.3731R/abstract};
(10)~\citet{https://ui.adsabs.harvard.edu/abs/2019ApJ...883...79D/abstract};
(11)~\citet{https://ui.adsabs.harvard.edu/abs/2022ApJS..259...62I/abstract};
(12)~Kepler Objects of Interest Catalog \citep{thompson2018};
(13)~\citet{yee2018};
(14)~\citet{bakos2010};
(15)~\citet{https://ui.adsabs.harvard.edu/abs/2022AJ....163..228P/abstract};
(16)~~TESS Objects of Interest Catalog Catalog \citep{guerrero2021};
(17)~\citet{https://ui.adsabs.harvard.edu/abs/2018A&A...613A..41M/abstract};
(18)~\citet{https://ui.adsabs.harvard.edu/abs/2019MNRAS.486.2290O/abstract};
(19)~\citet{https://ui.adsabs.harvard.edu/abs/2009ApJ...706..785H/abstract};
(20)~\citet{https://ui.adsabs.harvard.edu/abs/2012ApJ...749..134H/abstract};
(21)~\citet{bonomo2017};
(22)~\citet{https://ui.adsabs.harvard.edu/abs/2022arXiv221113761F/abstract};
(23)~\citet{https://ui.adsabs.harvard.edu/abs/2011ApJ...726...52H/abstract};
(24)~\citet{https://ui.adsabs.harvard.edu/abs/2015MNRAS.451.4060S/abstract};
(25)~\citet{https://ui.adsabs.harvard.edu/abs/2017A&A...601A..53E/abstract};
(26)~\citet{https://ui.adsabs.harvard.edu/abs/2010arXiv1008.3388B/abstract};
(27)~\citet{https://ui.adsabs.harvard.edu/abs/2011ApJ...728..138H/abstract};
(28)~\citet{https://ui.adsabs.harvard.edu/abs/2017AJ....153..136S/abstract};
(29)~\citet{https://ui.adsabs.harvard.edu/abs/2011AJ....141..179C/abstract};
(30)~\citet{torres2008};
(31)~\citet{https://ui.adsabs.harvard.edu/abs/2021AJ....161...64L/abstract};
(32)~\citet{mohler2013};
(33)~\citet{https://ui.adsabs.harvard.edu/abs/2017MNRAS.468..835B/abstract};
(34)~\citet{https://ui.adsabs.harvard.edu/abs/2015AJ....149..166H/abstract};
(35)~\citet{https://ui.adsabs.harvard.edu/abs/2020AJ....159..267B/abstract};
(36)~\citet{https://ui.adsabs.harvard.edu/abs/2019PASP..131k5003A/abstract};
(37)~\citet{https://ui.adsabs.harvard.edu/abs/2021ApJS..255....8R/abstract};
(38)~\citet{https://ui.adsabs.harvard.edu/abs/2023A&A...673A...4B/abstract};
(39)~\citet{https://ui.adsabs.harvard.edu/abs/2021AJ....161..117S/abstract};
(40)~\citet{https://ui.adsabs.harvard.edu/abs/2023AJ....165..134O/abstract};
(41)~\citet{https://ui.adsabs.harvard.edu/abs/2020MNRAS.tmp.1860D/abstract};
(42)~\citet{https://ui.adsabs.harvard.edu/abs/2021arXiv211013069B/abstract};
(43)~\citet{https://ui.adsabs.harvard.edu/abs/2021AJ....162..118E/abstract};
(44)~\citet{https://ui.adsabs.harvard.edu/abs/2020A&A...639A..76N/abstract};
(45)~\citet{https://ui.adsabs.harvard.edu/abs/2021AJ....162..176P/abstract};
(46)~\citet{https://ui.adsabs.harvard.edu/abs/2019ApJS..244...11K/abstract};
(47)~\citet{https://ui.adsabs.harvard.edu/abs/2020AJ....159..120L/abstract};
(48)~\citet{https://ui.adsabs.harvard.edu/abs/2022ApJS..258...40K/abstract};
(49)~\citet{https://ui.adsabs.harvard.edu/abs/2020AJ....160..192S/abstract};
(50)~\citet{https://ui.adsabs.harvard.edu/abs/2016ApJ...818...46M/abstract};
(51)~\citet{https://ui.adsabs.harvard.edu/abs/2016ApJS..226....7C/abstract};
(52)~\citet{https://ui.adsabs.harvard.edu/abs/2018AJ....155..136M/abstract};
(53)~\citet{https://ui.adsabs.harvard.edu/abs/2016ApJ...824...55S/abstract};
(54)~\citet{https://ui.adsabs.harvard.edu/abs/2019AcA....69..135S/abstract};
(55)~\citet{https://ui.adsabs.harvard.edu/abs/2018PASP..130g4401B/abstract};
(56)~\citet{https://ui.adsabs.harvard.edu/abs/2019RAA....19...41G/abstract};
(57)~\citet{https://ui.adsabs.harvard.edu/abs/2019AJ....158..138S/abstract};
(58)~\citet{sun2019};
(59)~\citet{https://ui.adsabs.harvard.edu/abs/2022Sci...377.1211L/abstract};
(60)~\citet{https://ui.adsabs.harvard.edu/abs/2021A&A...653A..41D/abstract};
(61)~\citet{https://ui.adsabs.harvard.edu/abs/2019A&A...629A.111C/abstract};
(62)~\citet{https://ui.adsabs.harvard.edu/abs/2020A&A...642A.121L/abstract};
(63)~\citet{https://ui.adsabs.harvard.edu/abs/2019AJ....157...32M/abstract};
(64)~\citet{https://ui.adsabs.harvard.edu/abs/2020A&A...644A.127D/abstract};
(65)~\citet{https://ui.adsabs.harvard.edu/abs/2018MNRAS.475.4467B/abstract};
(66)~\citet{https://ui.adsabs.harvard.edu/abs/2019A&A...625A.142E/abstract};
(67)~\citet{https://ui.adsabs.harvard.edu/abs/2017AJ....153..200A/abstract};
(68)~\citet{https://ui.adsabs.harvard.edu/abs/2018AJ....155...52A/abstract};
(69)~\citet{https://ui.adsabs.harvard.edu/abs/2019AJ....157..224A/abstract};
(70)~\citet{https://ui.adsabs.harvard.edu/abs/2020ApJ...892L...7H/abstract};
(71)~\citet{dong2022};
(72)~\citet{subjak2022};
(73)~\citet{https://ui.adsabs.harvard.edu/abs/2021AJ....162..144A/abstract};
(74)~\citet{https://ui.adsabs.harvard.edu/abs/2020ApJ...899...29K/abstract};
(75)~\citet{https://ui.adsabs.harvard.edu/abs/2022AJ....163..133E/abstract};
(76)~\citet{https://ui.adsabs.harvard.edu/abs/2023MNRAS.tmp..527M/abstract};
(77)~\citet{https://ui.adsabs.harvard.edu/abs/2020AJ....160..147C/abstract};
(78)~\citet{https://ui.adsabs.harvard.edu/abs/2022arXiv220303194O/abstract};
(79)~\citet{https://ui.adsabs.harvard.edu/abs/2021AJ....162...54H/abstract};
(80)~\citet{https://ui.adsabs.harvard.edu/abs/2021AJ....162..283T/abstract};
(81)~\citet{https://ui.adsabs.harvard.edu/abs/2021MNRAS.507.2154V/abstract};
(82)~\citet{https://ui.adsabs.harvard.edu/abs/2019NatAs...3.1099G/abstract};
(83)~\citet{https://ui.adsabs.harvard.edu/abs/2023AJ....165...84M/abstract};
(84)~\citet{https://ui.adsabs.harvard.edu/abs/2023arXiv230210008H/abstract};
(85)~\citet{https://ui.adsabs.harvard.edu/abs/2023AJ....166..163H/abstract};
(86)~\citet{https://ui.adsabs.harvard.edu/abs/2022AJ....164...81K/abstract};
(87)~\citet{https://ui.adsabs.harvard.edu/abs/2023AJ....166...44P/abstract};
(88)~\citet{almenara2022};
(89)~\citet{https://ui.adsabs.harvard.edu/abs/2023ApJ...954L..15M/abstract};
(90)~\citet{https://ui.adsabs.harvard.edu/abs/2023MNRAS.525L..98T/abstract};
(91)~\citet{https://ui.adsabs.harvard.edu/abs/2023arXiv230414703K/abstract};
(92)~\citet{https://ui.adsabs.harvard.edu/abs/2023AJ....166...30C/abstract};
(93)~\citet{https://ui.adsabs.harvard.edu/abs/2021A&A...653A..60M/abstract};
(94)~\citet{https://ui.adsabs.harvard.edu/abs/2023arXiv230103617G/abstract};
(95)~\citet{https://ui.adsabs.harvard.edu/abs/2023MNRAS.tmp..458H/abstract};
(96)~\citet{https://ui.adsabs.harvard.edu/abs/2015MNRAS.450.3101B/abstract};
(97)~\citet{https://ui.adsabs.harvard.edu/abs/2016PASP..128b4402S/abstract};
(98)~\citet{https://ui.adsabs.harvard.edu/abs/2009MNRAS.392.1585C/abstract};
(99)~\citet{anderson2017};
(100)~\citet{mocnik2017};
(101)~\citet{dai2017};
(102)~\citet{https://ui.adsabs.harvard.edu/abs/2009A&A...502..395W/abstract};
(103)~\citet{https://ui.adsabs.harvard.edu/abs/2022arXiv220502501H/abstract};
(104)~\citet{https://ui.adsabs.harvard.edu/abs/2017MNRAS.465.3693H/abstract};
(105)~\citet{https://ui.adsabs.harvard.edu/abs/2019MNRAS.482.1379H/abstract};
(106)~\citet{https://ui.adsabs.harvard.edu/abs/2018A&A...610A..63D/abstract};
(107)~\citet{https://ui.adsabs.harvard.edu/abs/2019MNRAS.482..301L/abstract};
(108)~\citet{https://ui.adsabs.harvard.edu/abs/2019MNRAS.485.5790T/abstract};
(109)~\citet{https://ui.adsabs.harvard.edu/abs/2011arXiv1104.2823H/abstract};
(110)~\citet{https://ui.adsabs.harvard.edu/abs/2011arXiv1105.3179A/abstract};
(111)~\citet{https://ui.adsabs.harvard.edu/abs/2014MNRAS.445.1114A/abstract};
(112)~\citet{https://ui.adsabs.harvard.edu/abs/2015MNRAS.450.2279T/abstract};
(113)~\citet{triaud2013};
(114)~\citet{https://ui.adsabs.harvard.edu/abs/2023MNRAS.tmpL..77M/abstract};
(115)~\citet{polanski2021}.
    }
}
\twocolumn

\section{Corner plot \label{appendix:cornerplot}}

\begin{figure*}
   \centering
   \includegraphics[width=\textwidth]{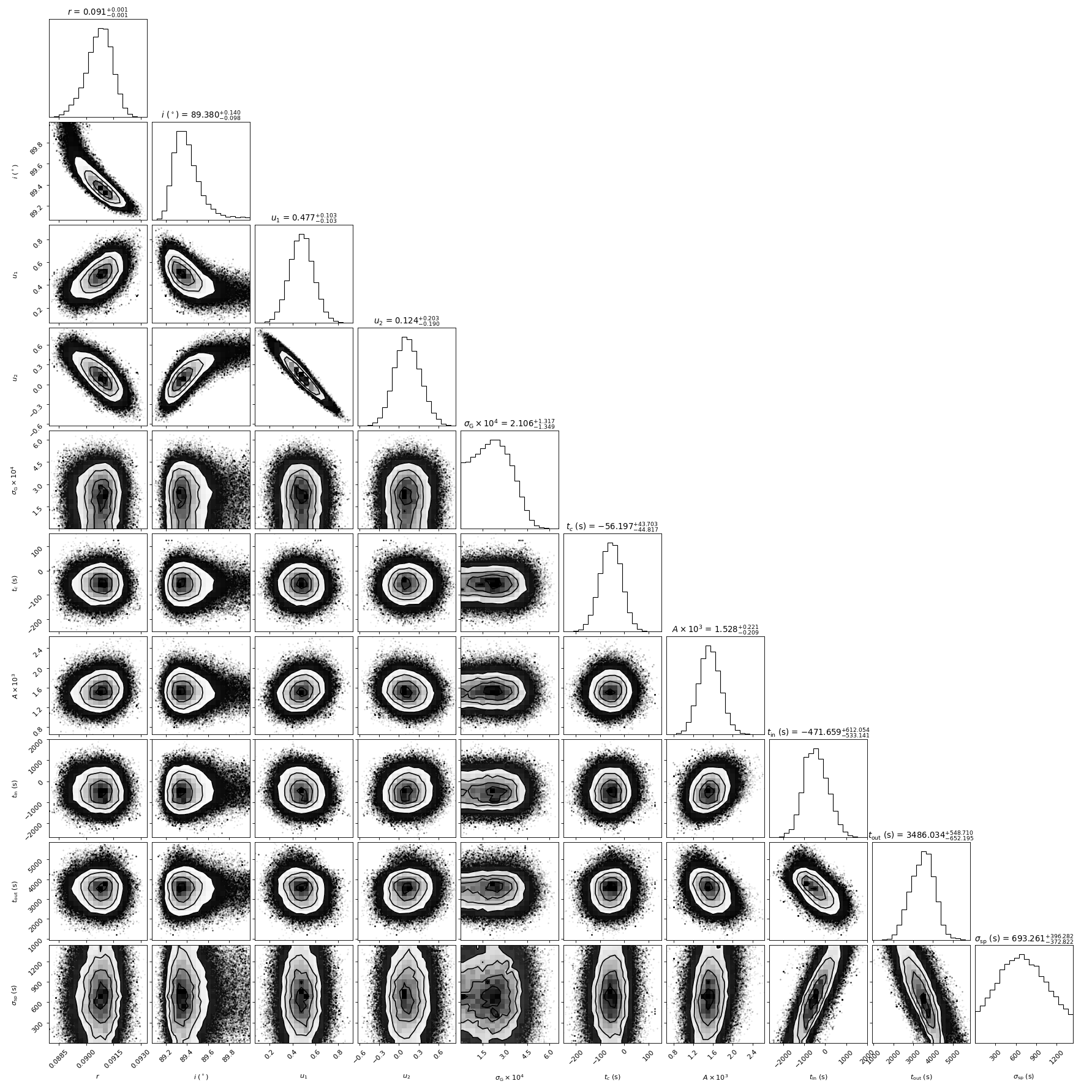}
   \caption{Corner plots of modelled parameters obtained from sampling the transit light curves of TOI-1268~b. The first four columns from the left ($r$, $i$, $u_1$, $u_2$) are parameters are uniform for the whole modelled segment of transit light curves. The fifth column gives the variance of the excess white noise in the observations. The last five columns ($t_{\rm c}$, $A$, $t_{\rm in}$, $t_{\rm out}$, $\sigma_{\rm sp}$) show parameters obtained from sampling transit~3 of the planet, which had decisive evidence for one spot occultation event.
         \label{fig:cornerplot}}
    \end{figure*}

In this appendix, we show the corner plot from sampling the transit light curves of TOI-1268~b. Fig.~\ref{fig:cornerplot} shows results for the first segment of TOI-1268~b light curves, containing one transit with decisive evidence for a spot occultation.

\section{Case study: Jupiter's transit in front of the Sun \label{appendix:jupiter_transit}}

To illustrate how the identified signals in our sample may correspond to actual features on stellar surfaces, we modelled the transit of Jupiter in front of the Sun as would be observed from outside the Solar System by a TESS-like space telescope.

We accessed the continuum image of the solar disk from the Helioseismic and Magnetic Imager (HMI) of NASA's Solar Dynamic Observatory\footnote{\url{https://sdo.gsfc.nasa.gov/data/aiahmi/}} (SDO) taken on October 23, 2014 at 13:15 UTC. HMI uses a narrow filter ($\Delta \lambda = 75$~m\AA) around $\lambda = 6173$~\AA \  (Fig.~\ref{fig:jupiter_transit}).

We refer to each pixel of the image as a solar surface element. We simulated the transit light curves of Jupiter as viewed from outside the Solar System with the following steps:

1.~We calculated the position of the centre of Jupiter's disk in 120~s intervals, converted to the pixel scale of the SDO image.

2.~Within a circle of radius $r_{\mathrm{Jup}}$ (scaled based on the size of the Sun) we set the flux of each solar surface element to zero. (We assumed that Jupiter has a fully opaque disk).

3.~We summed the flux of the solar surface elements to get the observed brightness of the Sun, and added some Gaussian noise to mimic the observations from TESS.

The spot signal is heavily dependent on the inclination of the transiting planet. The inclination of the planet's orbit sets which stellar latitude is transited by the planet. To investigate how the spot signal changes with for different transited latitudes, we simulated 20 transit light curves for Jupiter, changing the orbital inclination between 89.994 and 89.974 degrees by equal intervals. These inclinations cover the latitude range between 6.4 and 30.6 degrees.

To create light curves with no spot occultations, we modelled the transits at similar latitudes in front of the northern hemisphere of the Sun, which when the SDO image was taken did not show any detectable spots.

   The strongest spot detection statistics were achieved when Jupiter transited  $-17.5 ^\circ$ solar latitude. The likelihood ratio of the spot eclipse and the reference models was $\log r = 41.01$, while the Bayes factor in favour of the spot model was $\log_{10} B_{1,0} = 12.91$, which corresponds to a decisive evidence in favour of the model with one eclipsed spot.

   For the latitude with the highest Bayes-factor, the spot radius is $r_{\mathrm{min}} = 1.68_{-1.01}^{+2.06} \, ^\circ$ when the transit and end times are calculated without the $\sigma_{\mathrm{spot}}$ parameter and $r_{\mathrm{min}} = 5.36_{-1.53}^{+3.05} \, ^\circ$ when the $\sigma_{\mathrm{spot}}$ is included. The latter of these values  corresponds closely to the size of the active region when inspected visually (Fig.~\ref{fig:jupiter_transit}), which motivates the choice of $t_{\mathrm{in}}-\sigma_{\mathrm{spot}}$ and $t_{\mathrm{out}}+\sigma_{\mathrm{spot}}$ as the beginning and the end of the spot eclipse.

   We estimate the contrast of the active region to be $c = 0.15 \pm 0.01$, which with an effective temperature of the photosphere of $T_{\mathrm{eff}} = 5780$~K yields a spot temperature of $T_{\mathrm{spot}} = 5560 \pm 15$~K. Although the temperature of sunspots fall between 3000~K and 4500~K, we hold this number to be realistic, as the amplitude of the spot eclipse corresponds to the mean contrast of the active region, of which the cool umbrae are just a small fraction.
   
   We claim, therefore, that the contrast values obtained in our work can serve to calculate the mean temperature of active regions, or an upper limit for the temperature of individual spots. The inferred minimal radii $r_{\rm min}$ of the eclipsed regions should closely correspond to the angular width of the active region in the band in which the planet transited in front of it, and can be used as a lower limit to estimate the sizes of active regions.

\section{Spot occultations by HAT-P-11 b}

In Table~\ref{tab:hatp11bspots}, we provide detailed information about the candidate spot eclipses of HAT-P-11~b. In Table~\ref{tab:sap_comparison11}, we compare the amplitude of brightness variations $A_{\rm phot}$ in the Kepler observations of HAT-P-11 around each spot occultation event and the maximum possible flux change $\Delta F_{\rm spot}$ the spot candidate can cause on the star.

\onecolumn{
\begin{longtable}{lrrrrrrrr}
\caption{\label{tab:hatp11bspots} Significance and properties of candidate spot eclipse events identified in this analysis for HAT-P-11~b.}\\
\hline\hline
\multicolumn{1}{c}{Transit index} & \multicolumn{1}{c}{\begin{tabular}[c]{@{}c@{}}$t_c$\\ (BJD-2450000)\end{tabular}} & \multicolumn{1}{c}{$n_{\rm spot}$} & \multicolumn{1}{c}{$\log l$} & \multicolumn{1}{c}{$\log_{10} B$} & \multicolumn{1}{c}{\begin{tabular}[c]{@{}c@{}}$t_{\rm len}$\\ (s)\end{tabular}} & \multicolumn{1}{c}{\begin{tabular}[c]{@{}c@{}}$t_{\rm mid}$\\ (s)\end{tabular}} & \multicolumn{1}{c}{\begin{tabular}[c]{@{}c@{}}$r_{\mathrm{min}}$\\ ($^\circ$)\end{tabular}} & \multicolumn{1}{c}{$c$} \\
\hline
\endfirsthead
\caption{continued.}\\
\hline\hline
\multicolumn{1}{c}{Transit index} & \multicolumn{1}{c}{\begin{tabular}[c]{@{}c@{}}$t_c$\\ (BJD-2450000)\end{tabular}} & \multicolumn{1}{c}{$n_{\rm spot}$} & \multicolumn{1}{c}{$\log l$} & \multicolumn{1}{c}{$\log_{10} B$} & \multicolumn{1}{c}{\begin{tabular}[c]{@{}c@{}}$t_{\rm len}$\\ (s)\end{tabular}} & \multicolumn{1}{c}{\begin{tabular}[c]{@{}c@{}}$t_{\rm mid}$\\ (s)\end{tabular}} & \multicolumn{1}{c}{\begin{tabular}[c]{@{}c@{}}$r_{\mathrm{min}}$\\ ($^\circ$)\end{tabular}} & \multicolumn{1}{c}{$c$} \\
\hline
\endhead
\hline
\endfoot
$-523$ & $5075.12$ & 1 &  81.45 &  30.10 &$506_{-72}^{+104}$& $-414_{-11}^{+9}$ & $2.2_{-0.3}^{+0.1}$ &$0.257_{-0.026}^{+0.021}$ \\ 
$-522$ & $5080.01$ & 1 & 137.18 &  54.30 &$2499_{-173}^{+142}$& $-806_{-17}^{+29}$ & $8.6_{-0.7}^{+0.5}$ & $0.134_{-0.019}^{+0.008}$ \\
$-519$ & $5094.67$ & 1 &  98.12 &  37.34 &$299_{-147}^{+174}$& $2093_{-16}^{+16}$ & $1.9_{-0.2}^{+0.3}$ &$0.648_{-0.062}^{+0.059}$ \\ 
$-518$ & $5099.56$ & 1 &  59.97 &  20.77 &$1453_{-464}^{+498}$& $-590_{-53}^{+53}$ & $6.5_{-0.7}^{+0.7}$ & $0.124_{-0.010}^{+0.010}$ \\
$-516$ & $5109.33$ & 1 & 128.44 &  50.53 &$530_{-109}^{+114}$& $1559_{-15}^{+16}$ & $2.1_{-0.2}^{+0.2}$ & $0.216_{-0.016}^{+0.017}$ \\
$-514$ & $5119.11$ & 2 &  30.21 &   7.85 &$782_{-200}^{+276}$& $-1141_{-24}^{+25}$ & $3.0_{-0.5}^{+0.4}$ & $0.101_{-0.011}^{+0.012}$ \\
$-514$ & $5119.11$ & 2 &  30.21 &   7.85 &$365_{-119}^{+147}$& $910_{-21}^{+24}$ & $1.6_{-0.2}^{+0.3}$ &$0.195_{-0.036}^{+0.034}$ \\ 
$-512$ & $5128.89$ & 1 &  63.13 &  22.14 &$1079_{-337}^{+191}$& $-309_{-18}^{+17}$ & $3.0_{-0.2}^{+0.2}$ & $0.099_{-0.011}^{+0.013}$ \\
$-511$ & $5133.77$ & 1 &  34.06 &   9.52 &$429_{-125}^{+131}$& $1335_{-15}^{+18}$ & $1.8_{-0.2}^{+0.2}$ &$0.131_{-0.019}^{+0.019}$ \\ 
$-510$ & $5138.66$ & 2 & 235.84 &  97.15 &$580_{-90}^{+131}$& $-1303_{-9}^{+10}$ & $2.5_{-0.1}^{+0.2}$ & $0.265_{-0.015}^{+0.009}$ \\
$-510$ & $5138.66$ & 2 & 235.84 &  97.15 &$730_{-214}^{+177}$& $1709_{-15}^{+14}$ & $3.2_{-0.2}^{+0.2}$ & $0.168_{-0.009}^{+0.008}$ \\
$-509$ & $5143.55$ & 1 &  15.27 &   1.36 &$492_{-206}^{+221}$& $-1359_{-30}^{+22}$ & $2.2_{-0.3}^{+0.4}$ &$0.091_{-0.010}^{+0.012}$ \\ 
$-508$ & $5148.44$ & 1 &  44.51 &  14.06 &$600_{-114}^{+153}$& $892_{-28}^{+25}$ & $3.0_{-0.4}^{+0.3}$ & $0.121_{-0.017}^{+0.012}$ \\
$-507$ & $5153.33$ & 1 &  59.33 &  20.52 &$1076_{-269}^{+194}$& $2183_{-25}^{+24}$ & $3.8_{-0.3}^{+0.3}$ & $0.162_{-0.015}^{+0.014}$ \\
$-505$ & $5163.10$ & 2 &  51.29 &  17.00 &$517_{-220}^{+216}$& $-253_{-37}^{+32}$ & $2.4_{-0.3}^{+0.3}$ & $0.106_{-0.017}^{+0.015}$ \\
$-505$ & $5163.10$ & 2 &  51.29 &  17.00 &$1306_{-426}^{+361}$& $1460_{-35}^{+34}$ & $3.2_{-0.5}^{+0.5}$ & $0.134_{-0.013}^{+0.015}$ \\
$-504$ & $5167.99$ & 2 & 123.44 &  48.34 &$355_{-161}^{+181}$& $-819_{-25}^{+31}$ & $2.2_{-0.2}^{+0.3}$ &$0.298_{-0.027}^{+0.040}$ \\ 
$-504$ & $5167.99$ & 2 & 123.44 &  48.34 &$855_{-130}^{+122}$& $1668_{-14}^{+10}$ & $3.5_{-0.2}^{+0.1}$ & $0.333_{-0.016}^{+0.017}$ \\
$-487$ & $5251.08$ & 1 &  29.92 &   7.75 &$497_{-168}^{+290}$& $1845_{-22}^{+11}$ & $2.6_{-0.3}^{+0.2}$ &$0.084_{-0.004}^{+0.007}$ \\ 
$-486$ & $5255.97$ & 1 & 114.49 &  44.45 &$699_{-360}^{+530}$& $1474_{-22}^{+20}$ & $3.8_{-0.2}^{+0.4}$ & $0.124_{-0.006}^{+0.006}$ \\
$-485$ & $5260.86$ & 1 &  64.60 &  22.78 &$205_{-144}^{+109}$& $-416_{-26}^{+17}$ & $1.7_{-0.4}^{+0.2}$ &$0.422_{-0.082}^{+0.050}$ \\ 
$-484$ & $5265.74$ & 1 & 185.59 &  75.33 &$315_{-92}^{+110}$& $587_{-6}^{+7}$ & $2.4_{-0.1}^{+0.1}$ &$0.688_{-0.021}^{+0.053}$ \\ 
$-483$ & $5270.63$ & 1 & 296.91 & 123.67 &$867_{-78}^{+103}$& $-92_{-7}^{+6}$ & $5.0_{-0.1}^{+0.1}$ & $0.254_{-0.006}^{+0.005}$ \\
$-481$ & $5280.41$ & 1 & 116.90 &  45.49 &$972_{-170}^{+107}$& $3170_{-17}^{+14}$ & $2.5_{-0.3}^{+0.2}$ & $0.154_{-0.013}^{+0.019}$ \\
$-480$ & $5285.30$ & 1 & 164.85 &  66.32 &$451_{-28}^{+41}$& $1098_{-14}^{+19}$ & $2.6_{-0.2}^{+0.2}$ &$0.272_{-0.018}^{+0.020}$ \\ 
$-479$ & $5290.18$ & 1 &  17.24 &   2.21 &$767_{-248}^{+167}$& $2028_{-26}^{+27}$ & $5.7_{-0.3}^{+0.3}$ & $0.113_{-0.017}^{+0.008}$ \\
$-478$ & $5295.07$ & 1 & 327.06 & 136.77 &$967_{-70}^{+151}$& $879_{-4}^{+10}$ & $2.6_{-0.2}^{+0.1}$ & $0.255_{-0.007}^{+0.004}$ \\
$-476$ & $5304.85$ & 2 &  23.24 &   4.84 &$385_{-43}^{+115}$& $-616_{-12}^{+24}$ & $2.5_{-0.2}^{+0.1}$ &$0.218_{-0.020}^{+0.021}$ \\ 
$-476$ & $5304.85$ & 2 &  23.24 &   4.84 &$577_{-153}^{+81}$& $1181_{-39}^{+28}$ & $2.0_{-0.2}^{+0.2}$ & $0.095_{-0.018}^{+0.004}$ \\
$-474$ & $5314.62$ & 2 &  44.16 &  13.91 &$917_{-193}^{+136}$& $1309_{-10}^{+16}$ & $2.6_{-0.1}^{+0.0}$ & $0.282_{-0.012}^{+0.015}$ \\
$-474$ & $5314.62$ & 2 &  44.16 &  13.91 &$1032_{-225}^{+177}$& $1646_{-37}^{+13}$ & $4.1_{-0.2}^{+0.1}$ & $0.161_{-0.004}^{+0.007}$ \\
$-472$ & $5324.40$ & 1 & 121.17 &  47.35 &$526_{-187}^{+263}$& $1539_{-12}^{+17}$ & $9.4_{-0.1}^{+0.1}$ & $0.117_{-0.002}^{+0.004}$ \\
$-471$ & $5329.29$ & 1 &  42.04 &  12.98 &$400_{-75}^{+173}$& $2155_{-30}^{+14}$ & $2.7_{-0.4}^{+0.2}$ &$0.233_{-0.023}^{+0.023}$ \\ 
$-470$ & $5334.17$ & 2 & 146.32 &  58.27 &$556_{-95}^{+95}$& $-588_{-11}^{+25}$ & $1.9_{-0.0}^{+0.1}$ & $0.230_{-0.004}^{+0.005}$ \\
$-470$ & $5334.17$ & 2 & 146.32 &  58.27 &$643_{-91}^{+58}$& $1254_{-7}^{+4}$ & $2.7_{-0.1}^{+0.1}$ & $0.166_{-0.005}^{+0.001}$ \\
$-469$ & $5339.06$ & 2 & 278.98 & 115.89 &$1429_{-140}^{+136}$& $-813_{-11}^{+6}$ & $2.4_{-0.1}^{+0.0}$ & $0.312_{-0.008}^{+0.005}$ \\
$-469$ & $5339.06$ & 2 & 278.98 & 115.89 &$601_{-117}^{+311}$& $2544_{-13}^{+20}$ & $7.1_{-0.2}^{+0.2}$ & $0.100_{-0.005}^{+0.002}$ \\
$-468$ & $5343.95$ & 1 & 313.09 & 130.70 &$1492_{-225}^{+222}$& $1552_{-6}^{+9}$ & $4.0_{-0.2}^{+0.5}$ & $0.242_{-0.018}^{+0.023}$ \\
$-466$ & $5353.73$ & 1 & 147.42 &  58.75 &$769_{-157}^{+142}$& $1603_{-12}^{+14}$ & $6.1_{-0.1}^{+0.2}$ & $0.152_{-0.010}^{+0.001}$ \\
$-465$ & $5358.61$ & 1 &  46.96 &  15.12 &$277_{-59}^{+76}$& $1497_{-30}^{+27}$ & $2.0_{-0.3}^{+0.3}$ &$0.290_{-0.048}^{+0.045}$ \\ 
$-464$ & $5363.50$ & 1 &  46.24 &  14.81 &$951_{-113}^{+89}$& $-466_{-15}^{+18}$ & $1.7_{-0.1}^{+0.4}$ & $0.096_{-0.014}^{+0.011}$ \\
$-463$ & $5368.39$ & 2 & 143.72 &  57.14 &$399_{-172}^{+192}$& $-1626_{-10}^{+22}$ & $2.5_{-0.3}^{+0.2}$ &$0.334_{-0.022}^{+0.020}$ \\ 
$-463$ & $5368.39$ & 2 & 143.72 &  57.14 &$2063_{-238}^{+212}$& $2150_{-98}^{+24}$ & $8.1_{-0.4}^{+0.5}$ & $0.175_{-0.004}^{+0.004}$ \\
$-462$ & $5373.28$ & 1 & 130.69 &  51.48 &$535_{-225}^{+291}$& $1419_{-16}^{+17}$ & $2.3_{-0.2}^{+0.2}$ & $0.224_{-0.011}^{+0.015}$ \\
$-460$ & $5383.05$ & 1 & 267.57 & 110.95 &$636_{-114}^{+171}$& $1583_{-12}^{+15}$ & $5.2_{-0.2}^{+0.2}$ & $0.209_{-0.008}^{+0.009}$ \\
$-456$ & $5402.60$ & 2 &  29.44 &   7.54 &$515_{-167}^{+302}$& $1394_{-16}^{+34}$ & $1.8_{-0.2}^{+0.4}$ & $0.109_{-0.010}^{+0.006}$ \\
$-456$ & $5402.60$ & 2 &  29.44 &   7.54 &$536_{-109}^{+263}$& $2039_{-14}^{+19}$ & $2.9_{-0.3}^{+0.2}$ & $0.223_{-0.006}^{+0.010}$ \\
$-455$ & $5407.49$ & 1 &  89.64 &  33.66 &$1113_{-339}^{+357}$& $-951_{-24}^{+32}$ & $4.0_{-0.3}^{+0.4}$ & $0.121_{-0.009}^{+0.010}$ \\
$-453$ & $5417.27$ & 2 &  53.39 &  17.94 &$1393_{-337}^{+380}$& $-700_{-47}^{+64}$ & $5.5_{-0.6}^{+0.7}$ & $0.082_{-0.010}^{+0.007}$ \\
$-453$ & $5417.27$ & 2 &  53.39 &  17.94 &$1443_{-507}^{+344}$& $1906_{-47}^{+22}$ & $7.0_{-0.3}^{+0.2}$ & $0.094_{-0.006}^{+0.006}$ \\
$-451$ & $5427.04$ & 2 &  23.78 &   5.05 &$2090_{-850}^{+1273}$& $381_{-202}^{+131}$ & $13.3_{-1.7}^{+1.5}$ & $0.040_{-0.005}^{+0.005}$ \\
$-451$ & $5427.04$ & 2 &  23.78 &   5.05 &$811_{-542}^{+942}$& $1634_{-28}^{+24}$ & $2.6_{-0.3}^{+0.5}$ & $0.095_{-0.015}^{+0.017}$ \\
$-447$ & $5446.59$ & 1 & 131.13 &  51.68 &$1216_{-249}^{+264}$& $1845_{-16}^{+27}$ & $5.6_{-0.4}^{+0.3}$ & $0.154_{-0.007}^{+0.011}$ \\
$-446$ & $5451.48$ & 1 &  55.68 &  18.91 &$564_{-204}^{+219}$& $-815_{-26}^{+29}$ & $2.1_{-0.3}^{+0.3}$ & $0.114_{-0.014}^{+0.015}$ \\
$-407$ & $5642.11$ & 1 &  72.03 &  26.01 &$1713_{-520}^{+535}$& $1428_{-48}^{+50}$ & $7.3_{-0.6}^{+0.6}$ & $0.114_{-0.010}^{+0.009}$ \\
$-406$ & $5646.99$ & 1 &  11.04 &  -0.48 &$551_{-105}^{+136}$& $1650_{-18}^{+16}$ & $2.5_{-0.2}^{+0.2}$ & $0.186_{-0.015}^{+0.015}$ \\
$-405$ & $5651.88$ & 1 &  49.71 &  16.31 &$1788_{-672}^{+796}$& $1780_{-680}^{+110}$ & $6.0_{-1.3}^{+5.3}$ & $0.090_{-0.038}^{+0.045}$ \\
$-385$ & $5749.64$ & 1 &  78.28 &  28.72 &$589_{-101}^{+165}$& $2089_{-29}^{+18}$ & $2.1_{-0.1}^{+0.2}$ & $0.166_{-0.018}^{+0.010}$ \\
$-384$ & $5754.52$ & 1 &  45.86 &  14.64 &$881_{-118}^{+150}$& $1631_{-11}^{+16}$ & $2.7_{-0.1}^{+0.2}$ & $0.256_{-0.011}^{+0.019}$ \\
$-383$ & $5759.41$ & 1 &  40.82 &  12.45 &$436_{-7}^{+16}$& $1944_{-43}^{+45}$ & $1.9_{-0.5}^{+0.7}$ &$0.155_{-0.045}^{+0.050}$ \\ 
$-379$ & $5778.96$ & 1 &  42.99 &  13.42 &$959_{-112}^{+102}$& $2264_{0}^{+36}$ & $4.3_{0.0}^{+0.1}$ & $0.176_{-0.004}^{+0.002}$ \\
$-378$ & $5783.85$ & 1 & 147.24 &  58.67 &$605_{-61}^{+86}$& $1728_{-17}^{+16}$ & $2.8_{-0.2}^{+0.2}$ & $0.181_{-0.014}^{+0.014}$ \\
$-377$ & $5788.74$ & 1 & 137.79 &  54.57 &$773_{-90}^{+113}$& $1578_{-23}^{+21}$ & $3.3_{-0.3}^{+0.3}$ & $0.222_{-0.015}^{+0.016}$ \\
$-376$ & $5793.63$ & 1 & 144.49 &  57.48 &$579_{-131}^{+126}$& $1487_{-10}^{+11}$ & $1.9_{-0.2}^{+0.2}$ & $0.259_{-0.018}^{+0.019}$ \\
$-373$ & $5808.29$ & 1 &  70.61 &  25.39 &$1226_{-306}^{+155}$& $2059_{-19}^{+22}$ & $3.3_{-0.3}^{+0.3}$ & $0.224_{-0.015}^{+0.016}$ \\
$-371$ & $5818.07$ & 1 & 111.47 &  43.14 &$636_{-108}^{+149}$& $1240_{-22}^{+22}$ & $4.1_{-0.3}^{+0.2}$ & $0.140_{-0.010}^{+0.010}$ \\
$-370$ & $5822.95$ & 1 & 267.18 & 110.78 &$634_{-89}^{+152}$& $1335_{-7}^{+7}$ & $2.5_{-0.1}^{+0.1}$ & $0.434_{-0.010}^{+0.014}$ \\
$-369$ & $5827.84$ & 2 &  33.57 &   9.31 &$1254_{-265}^{+245}$& $-42_{-26}^{+29}$ & $2.4_{-0.3}^{+0.4}$ & $0.083_{-0.011}^{+0.008}$ \\
$-369$ & $5827.84$ & 2 &  33.57 &   9.31 &$4448_{-236}^{+262}$& $1736_{-18}^{+18}$ & $5.7_{-0.2}^{+0.2}$ & $0.121_{-0.010}^{+0.009}$ \\
$-347$ & $5935.37$ & 1 &  65.98 &  23.38 &$1169_{-344}^{+341}$& $1446_{-31}^{+25}$ & $9.7_{-0.3}^{+0.3}$ & $0.170_{-0.009}^{+0.011}$ \\
$-341$ & $5964.70$ & 1 &  70.47 &  25.33 &$2512_{-1434}^{+1390}$& $1330_{-42}^{+35}$ & $6.3_{-0.5}^{+0.6}$ & $0.173_{-0.011}^{+0.014}$ \\
$-332$ & $6008.69$ & 2 &  64.75 &  22.85 &$648_{-215}^{+264}$& $-62_{-47}^{+48}$ & $3.1_{-0.4}^{+0.4}$ & $0.091_{-0.007}^{+0.009}$ \\
$-332$ & $6008.69$ & 2 &  64.75 &  22.85 &$505_{-194}^{+197}$& $2610_{-15}^{+33}$ & $2.4_{-0.3}^{+0.4}$ &$0.161_{-0.030}^{+0.028}$ \\ 
$-331$ & $6013.58$ & 1 &  47.89 &  15.53 &$3142_{-1572}^{+631}$& $-870_{-27}^{+32}$ & $2.6_{-0.3}^{+0.5}$ & $0.135_{-0.024}^{+0.013}$ \\
$-309$ & $6121.11$ & 1 & 235.95 &  97.22 &$624_{-148}^{+191}$& $2041_{-12}^{+16}$ & $6.8_{-0.2}^{+0.2}$ & $0.191_{-0.006}^{+0.006}$ \\
$-306$ & $6135.77$ & 1 &  15.47 &   1.45 &$342_{-124}^{+304}$& $1350_{-16}^{+10}$ & $1.6_{-0.1}^{+0.2}$ &$0.188_{-0.032}^{+0.027}$ \\ 
$-305$ & $6140.66$ & 1 &  35.08 &   9.96 &$576_{-337}^{+384}$& $189_{-20}^{+28}$ & $2.3_{-0.3}^{+0.3}$ & $0.085_{-0.010}^{+0.013}$ \\
$-302$ & $6155.32$ & 1 &  48.06 &  15.60 &$2294_{-336}^{+248}$& $2114_{-24}^{+26}$ & $11.7_{-0.4}^{+0.3}$ & $0.094_{-0.004}^{+0.006}$ \\
$-296$ & $6184.65$ & 2 &  88.48 &  33.15 &$702_{-157}^{+210}$& $-271_{-24}^{+33}$ & $2.7_{-0.3}^{+0.4}$ & $0.184_{-0.020}^{+0.018}$ \\
$-296$ & $6184.65$ & 2 &  88.48 &  33.15 &$588_{-101}^{+135}$& $2443_{-18}^{+18}$ & $2.7_{-0.2}^{+0.2}$ & $0.361_{-0.028}^{+0.023}$ \\
$-271$ & $6306.85$ & 1 &  14.56 &   1.05 &$586_{-179}^{+166}$& $2797_{-25}^{+28}$ & $2.9_{-0.3}^{+0.3}$ & $0.119_{-0.010}^{+0.012}$ \\
$-261$ & $6355.72$ & 2 &  20.45 &   3.64 &$477_{-178}^{+138}$& $-1615_{-19}^{+27}$ & $2.0_{-0.2}^{+0.2}$ &$0.162_{-0.015}^{+0.024}$ \\ 
$-261$ & $6355.72$ & 2 &  20.45 &   3.64 &$385_{-162}^{+192}$& $2278_{-36}^{+19}$ & $1.9_{-0.2}^{+0.3}$ &$0.198_{-0.037}^{+0.036}$ \\ 
$-254$ & $6389.94$ & 2 &  21.86 &   4.22 &$791_{-352}^{+299}$& $-832_{-67}^{+66}$ & $2.8_{-0.7}^{+0.8}$ & $0.099_{-0.019}^{+0.025}$ \\
$-254$ & $6389.94$ & 2 &  21.86 &   4.22 &$524_{-153}^{+287}$& $2462_{-38}^{+38}$ & $2.8_{-0.6}^{+0.7}$ & $0.135_{-0.026}^{+0.028}$ \\
$-253$ & $6394.83$ & 1 & 113.99 &  44.23 &$2155_{-313}^{+314}$& $-1076_{-24}^{+25}$ & $10.3_{-0.2}^{+0.3}$ & $0.130_{-0.007}^{+0.006}$ \\
$-252$ & $6399.71$ & 1 & 103.30 &  39.62 &$679_{-119}^{+135}$& $1058_{-15}^{+15}$ & $2.4_{-0.2}^{+0.2}$ & $0.257_{-0.013}^{+0.018}$ \\
\end{longtable}
}

\begin{longtable}{rrrr}
\caption{\label{tab:sap_comparison11} Expected maximum flux change from co-rotating starspots $\Delta F_{\rm spot}$ compared to the amplitude of the rotational modulation observed around the transit $A_{\rm phot}$ for HAT-P-11. The data column tells whether amplitudes were derived from SAP or PDC light curves.}\\
\hline\hline
\multicolumn{1}{c}{Transit} & \multicolumn{1}{c}{Data} & \multicolumn{1}{c}{$A_{\rm phot}$ ($\times 10^5$)} & \multicolumn{1}{c}{$\Delta F_{\rm spot}$ ($\times 10^5$)}  \\
\hline
\endfirsthead
\caption{continued.}\\
\hline\hline
\multicolumn{1}{c}{Transit} & \multicolumn{1}{c}{Data} & \multicolumn{1}{c}{$A_{\rm phot}/10^{-5}$} & \multicolumn{1}{c}{$\Delta F_{\rm spot}/10^{-5}$}  \\
\hline
\endhead
\hline
\endfoot
$-523$ & SAP &1626&    $  19 \pm   18$ \\
$-522$ & SAP &1626&    $ 140 \pm   25$ \\
$-519$ & SAP &1626&    $  38 \pm   10$ \\
$-518$ & SAP &1626&    $  78 \pm   18$ \\
$-516$ & SAP &1024&    $  15 \pm    3$ \\
$-514$ & SAP &1175&    $  14 \pm    4$ \\
$-514$ & SAP &1175&    $   8 \pm    3$ \\
$-512$ & SAP &1214&    $  14 \pm    2$ \\
$-511$ & SAP &1231&    $   6 \pm    1$ \\
$-510$ & SAP &1231&    $  25 \pm    3$ \\
$-510$ & SAP &1231&    $  27 \pm    4$ \\
$-509$ & SAP &1231&    $   7 \pm    2$ \\
$-508$ & SAP &1698&    $  16 \pm    4$ \\
$-507$ & SAP &1711&    $  35 \pm    6$ \\
$-505$ & SAP &1699&    $   9 \pm    3$ \\
$-505$ & SAP &1699&    $  21 \pm    6$ \\
$-504$ & SAP &1699&    $  23 \pm    5$ \\
$-504$ & SAP &1699&    $  60 \pm    6$ \\
$-487$ & SAP &1128&    $   9 \pm    2$ \\
$-486$ & SAP &1221&    $  28 \pm    5$ \\
$-485$ & SAP &1221&    $  18 \pm    7$ \\
$-484$ & SAP &1221&    $  61 \pm    6$ \\
$-483$ & SAP &1221&    $  95 \pm    5$ \\
$-481$ & SAP &1123&    $  14 \pm    3$ \\
$-480$ & SAP &1118&    $  28 \pm    4$ \\
$-479$ & SAP &1020&    $  53 \pm    9$ \\
$-478$ & SAP &1044&    $  25 \pm    3$ \\
$-476$ & SAP &1264&    $  19 \pm    3$ \\
$-476$ & SAP &1264&    $   6 \pm    1$ \\
$-474$ & SAP &1368&    $  29 \pm    2$ \\
$-474$ & SAP &1368&    $  41 \pm    3$ \\
$-472$ & SAP &1369&    $ 161 \pm    3$ \\
$-471$ & SAP &1369&    $  24 \pm    6$ \\
$-470$ & SAP &1369&    $  13 \pm    1$ \\
$-470$ & SAP &1369&    $  19 \pm    1$ \\
$-469$ & SAP &1271&    $  27 \pm    1$ \\
$-469$ & SAP &1271&    $  77 \pm    4$ \\
$-468$ & SAP &1271&    $  62 \pm   11$ \\
$-466$ & SAP &1193&    $  84 \pm    5$ \\
$-465$ & SAP &1193&    $  17 \pm    5$ \\
$-464$ & SAP &1132&    $   5 \pm    1$ \\
$-463$ & SAP &643&    $  31 \pm    6$ \\
$-463$ & SAP &643&    $ 182 \pm   20$ \\
$-462$ & SAP &599&    $  18 \pm    3$ \\
$-460$ & SAP &480&    $  87 \pm    8$ \\
$-456$ & SAP &961&    $   6 \pm    2$ \\
$-456$ & SAP &961&    $  28 \pm    5$ \\
$-455$ & SAP &988&    $  30 \pm    6$ \\
$-453$ & SAP &534&    $  38 \pm   10$ \\
$-453$ & SAP &534&    $  66 \pm    9$ \\
$-451$ & SAP &540&    $ 100 \pm   26$ \\
$-451$ & SAP &540&    $  12 \pm    4$ \\
$-447$ & SAP &1385&    $  71 \pm   10$ \\
$-446$ & SAP &1385&    $   8 \pm    2$ \\
$-407$ & SAP &1293&    $  90 \pm   17$ \\
$-406$ & SAP &1287&    $  18 \pm    3$ \\
$-405$ & SAP &1895&    $  87 \pm   68$ \\
$-385$ & SAP &1810&    $  12 \pm    2$ \\
$-384$ & SAP &1810&    $  30 \pm    4$ \\
$-383$ & SAP &1875&    $  10 \pm    6$ \\
$-379$ & SAP &1865&    $  51 \pm    1$ \\
$-378$ & SAP &1865&    $  22 \pm    3$ \\
$-377$ & SAP &1865&    $  37 \pm    7$ \\
$-376$ & SAP &1451&    $  14 \pm    3$ \\
$-373$ & SAP &1011&    $  37 \pm    7$ \\
$-371$ & SAP &1011&    $  35 \pm    5$ \\
$-370$ & SAP &711&    $  41 \pm    3$ \\
$-369$ & SAP &556&    $   7 \pm    2$ \\
$-369$ & SAP &556&    $  59 \pm    6$ \\
$-347$ & SAP &1018&    $ 243 \pm   21$ \\
$-341$ & SAP &816&    $ 105 \pm   19$ \\
$-332$ & SAP &1480&    $  13 \pm    4$ \\
$-332$ & SAP &1480&    $  15 \pm    5$ \\
$-331$ & SAP &1480&    $  15 \pm    5$ \\
$-309$ & SAP &1110&    $ 134 \pm   11$ \\
$-306$ & SAP &1397&    $   7 \pm    2$ \\
$-305$ & SAP &1576&    $   7 \pm    2$ \\
$-302$ & SAP &1928&    $ 192 \pm   14$ \\
$-296$ & SAP &1994&    $  21 \pm    6$ \\
$-296$ & SAP &1994&    $  40 \pm    6$ \\
$-271$ & SAP &796&    $  16 \pm    3$ \\
$-261$ & SAP &1800&    $  10 \pm    2$ \\
$-261$ & SAP &1800&    $  12 \pm    4$ \\
$-254$ & SAP &988&    $  14 \pm    7$ \\
$-254$ & SAP &988&    $  18 \pm    9$ \\
$-253$ & SAP &1006&    $ 211 \pm   19$ \\
$-252$ & SAP &863&    $  23 \pm    4$ \\
\hline
\end{longtable}
   
\end{appendix}
\end{document}